\documentclass[journal]{IEEEtran}
\usepackage{pgfplots}
\usepackage{subcaption}

\usepackage{amssymb}
\usepackage{algorithm}
\usepackage{algpseudocode}
\usepackage{color}
\usepackage{epsf}
\usepackage{psfrag}
\usepackage{epsfig}
\usepackage{graphicx}
\usepackage{amsfonts}
\usepackage{textcomp}
\usepackage{comment}
\usepackage{footnote}
\usepackage{tablefootnote}
\usepackage[flushleft]{threeparttable}
\usepackage{paralist}
\usepackage{mdwlist}
\usepackage{amssymb}
\usepackage{amsmath}
\usepackage{url}
\usepackage{verbatim}
\usepackage{alltt}
\usepackage{balance}
\usepackage{multirow}
\usepackage{epstopdf}
\usepackage{textcomp}
\usepackage{bbm}
\usepackage{lipsum}
\usepackage{float}
\usepackage{booktabs}
\usepackage{slashbox}
\usepackage[table]{xcolor}
\usepackage{array}
\usepackage{boldline}
\usepackage{caption,setspace}
\usepackage{mathtools}
\usepackage{adjustbox}
\usepackage{upgreek}
\usepackage{ stmaryrd }

\usepackage{calligra}
\usepackage[T1]{fontenc}
\usepackage{ upgreek }
\usepackage{ mathrsfs }
\usepackage{ upgreek }

\usepackage{microtype}
\usepackage{lineno}

\definecolor{green1}{RGB}{0,175,0}
\definecolor{blue1}{RGB}{0,150,200}

\usepackage[colorlinks,urlcolor=black,linkcolor=blue,citecolor=blue]{hyperref}

\usepackage[normalem]{ulem}
\usepackage{cancel}
\usepackage{svg}

\usepackage{cite}
\usepackage{amsmath}
\usepackage{graphicx}

\usepackage{times}
\usepackage{latexsym}

\usepackage[T1]{fontenc}

\usepackage[utf8]{inputenc}

\usepackage{microtype}

\usepackage{inconsolata}

\usepackage{graphicx}
\usepackage{adjustbox}
\usepackage{multirow}
\usepackage{amsmath}
\usepackage{amsfonts}
\usepackage{xcolor}
\usepackage{mathrsfs}

\usepackage{xcolor}
\usepackage{pifont}
\usepackage{subcaption}
\usepackage{colortbl} 

\usepackage{tikz}
\pgfplotsset{compat=1.18}

\usepackage{array}
\newcommand{\cmark}{\ding{51}}
\newcommand{\xmark}{\ding{55}}

\ifCLASSINFOpdf
\else
\fi
\begin{document}


\title{Hierarchical Book Organization for Learning-Resource Discovery using Dual-Path Graph Convolutions}

\newcommand{\titleabbr}{HiGeMine}

\author{Suraj Kumar, \IEEEmembership{Graduate Student Member, IEEE,} 
Utsav Kumar Nareti, \IEEEmembership{Graduate Student Member, IEEE,} 
Soumi Chattopadhyay, \IEEEmembership{Senior Member, IEEE,}  
Chandranath Adak, \IEEEmembership{Senior Member, IEEE,} 
Prolay Mallick
\thanks{
S. Kumar, S. Chattopadhyay, and P. Mallick are with the Dept. of CSE, Indian Institute of Technology Indore, Madhya Pradesh 453552, India.
U. K. Nareti, and C. Adak are with the Dept. of CSE, Indian Institute of Technology Patna, Bihar 801106, India. 
Corresponding authors: S. Chattopadhyay (soumi@iiti.ac.in), C. Adak (email: chandranath@iitp.ac.in)
}
}

\maketitle

\begin{abstract}
The growing availability of books and textual materials in digital learning environments necessitates reliable semantic organization to support scalable resource management and discovery.
However, existing book classification approaches typically formulate genre prediction as a flat classification problem, overlooking both the hierarchical organization of literary categories and the semantic discrepancy between authoritative book descriptions and subjective crowd-sourced reviews.
We propose {\titleabbr}, a hierarchical book classification framework for structured learning-resource organization that reformulates genre prediction as coarse-to-fine semantic reasoning over heterogeneous textual evidence.
{\titleabbr} first performs blurb-guided semantic refinement to preserve semantically consistent reviews while suppressing noisy and genre-irrelevant interpretations. It then performs semantic-role-separated hierarchical graph reasoning through independent propagation branches for blurbs and reviews, enabling separate modeling of narrative and interpretive semantics during hierarchical inference. A coarse-grained level-1 classifier first distinguishes fiction from non-fiction, followed by domain-specialized level-2 multi-label classifiers for fine-grained genre prediction. 
Furthermore, {\titleabbr} captures 
relationships among fine-grained resource categories 
through a structured label co-occurrence graph and genre-conditioned semantic representations. 
To facilitate systematic evaluation, we curate a new hierarchical multi-label Goodreads benchmark containing paired blurbs and reviews. 
Extensive experiments against hierarchical classifiers, graph-based approaches, pretrained language models, and large language models demonstrate the effectiveness of {\titleabbr} for reliable hierarchical book classification, thereby providing a scalable foundation for organizing and discovering textual resources in digital learning environments.
\end{abstract}


\section{Introduction}

The rapid growth of digital repositories and online learning environments has increased the need for reliable organization and discovery of textual resources \cite{tlt1}. Books constitute an important class of such resources, where genre information provides structured semantic metadata for indexing, navigation, recommendation, and resource discovery. Genre classification has been widely studied across media such as movies, music, and books \cite{zhang_wacv_2024movie,music_cnn_2024_mta,p2comic2023,cover_iapr_2024}. Unlike movies and music, which benefit from multimodal cues such as visuals and audio, book classification predominantly relies on textual evidence such as titles and blurbs, as book covers often provide limited discriminative information \cite{exan_ijdar_2023}. With the growth of digital bookstores and online catalogs, reliable genre characterization has become increasingly important for book success prediction, personalized recommendation, catalog organization, and audience targeting \cite{genre_success_2020_saci,rnn_review_2022_ijit,cover_title_ICISGT_2019}. In digital learning environments, fine-grained genre metadata can further facilitate scalable organization and discovery of textual learning resources.

Reliable book organization remains challenging due to hierarchical,
multi-label genres and heterogeneous textual evidence. 
Books may be fiction or non-fiction 
while exhibiting multiple fine-grained genres, yet most approaches adopt flat multi-class prediction
\cite{benchmarking_cover}. Moreover, publisher-authored blurbs provide concise,
relatively authoritative descriptions, whereas crowdsourced reviews offer
richer thematic and perceptual cues but are often noisy, subjective, and
sentiment-driven; label sparsity and inconsistencies in resources such as
Goodreads further complicate genre inference \cite{judgingbookcover}. 
Existing methods only partially address these challenges. Early approaches
explored visual and textual cover cues \cite{judgingbookcover}, while
title-based text was shown to perform comparably to multimodal representations
\cite{cover_title_ICISGT_2019}. Recent methods exploit metadata, titles, blurbs,
and reviews
\cite{genre_success_2020_saci,Title_ICCCS2019,desc_spanish_ieee_access_2023,review_2018_aspect_wicwi},
using representations from TF-IDF to Word2Vec, FastText, and GloVe
\cite{book_review_Portuguese_Language2022}. However, they do not jointly address
\textit{(i)} noisy crowdsourced evidence, 
\textit{(ii)} semantic discrepancies
between authoritative descriptions and subjective interpretations, 
and
\textit{(iii)} hierarchical, interdependent genre structures. 
Thus, direct
blurb-review fusion may introduce semantic interference, while flat prediction
limits fine-grained book organization and resource discovery.

To address these gaps, we propose {\titleabbr}, a hierarchical book organization framework that transforms heterogeneous textual evidence into structured fine-grained genre metadata. {\titleabbr} first performs blurb-guided zero-shot semantic refinement to retain reviews semantically consistent with the corresponding book description while suppressing noisy and genre-irrelevant interpretations. Rather than prematurely fusing the two sources, it employs a semantic-role-separated dual-path graph architecture that propagates blurbs and reviews through independent graph spaces, preserving their respective narrative and interpretive semantics while enabling weighted evidence integration. Hierarchical inference follows a coarse-to-fine organization: a level-1 classifier distinguishes fiction from non-fiction, followed by domain-specialized level-2 multi-label classifiers for fine-grained genre prediction. Genre-conditioned representations and a structured label co-occurrence graph further capture category-specific semantics and inter-genre dependencies. The resulting hierarchical characterization provides structured semantic metadata for scalable book organization and learning-resource discovery. 
Our main \textbf{contributions} are:

\noindent
\textit{\textbf{(i)}} We propose {\titleabbr}, a hierarchical book organization framework that reformulates literary genre prediction as a coarse-to-fine semantic reasoning problem by disentangling authoritative narrative evidence (blurbs) from subjective crowdsourced interpretations (reviews). To achieve this, {\titleabbr} introduces a semantic-role-separated dual-path graph architecture that models blurbs and reviews through independent heterogeneous graph propagation spaces, preserving modality-specific semantics while enabling weighted evidence integration during hierarchical inference.

\noindent
\textit{\textbf{(ii)}} We introduce a blurb-conditioned semantic refinement mechanism that uses authoritative narrative descriptions to contextualize and refine crowdsourced reviews before genre inference. The framework selectively preserves semantically aligned evidence while suppressing noisy, sentiment-dominated, and genre-irrelevant interpretations, improving robustness under heterogeneous textual signals.

\noindent
\textit{\textbf{(iii)}} We develop a genre-conditioned representation learning framework in which token embeddings are derived from their normalized distribution across genre categories rather than relying solely on static pretrained lexical representations. Combined with a structured label co-occurrence graph and adaptive label refinement, the framework jointly captures genre-aware token semantics, inter-label dependencies, and fine-grained hierarchical multi-label interactions.

\noindent
\textit{\textbf{(iv)}} We curate a new hierarchical multi-label Goodreads benchmark containing paired blurbs and reviews for fine-grained fiction and non-fiction genre prediction. Extensive experiments against hierarchical classifiers, graph-based methods, language models, and recent LLMs demonstrate the effectiveness of {\titleabbr} for reliable hierarchical book classification, providing structured semantic metadata for scalable organization and discovery of textual resources in digital learning environments.

\section{Related Work}


The growing availability of digital learning materials has increased the need
for effective resource organization and discovery. Structured metadata
supports the indexing and retrieval of learning resources, while recommender
systems facilitate access to learner-relevant content \cite{tlt1}. Books
constitute an important class of textual resources, where genre information
provides fine-grained semantic metadata that complements conventional
metadata for resource organization and discovery.
Digital literary platforms such as Amazon Kindle, Google Books, and Goodreads
rely on genre information for indexing, recommendation, and discovery
\cite{likability_2018_cemnlp,cover_iapr_2024}. However, genre characterization
is challenging due to overlapping genres \cite{desc_2017_ICMMI,p2comic2023}
and subjective, inconsistent crowd-sourced annotations. Despite the natural
hierarchical and overlapping structure of book genres, most approaches adopt
flat or multiclass prediction
\cite{judgingbookcover,book_chapters_2018_ICCL,benchmarking_cover,desc_spanish_ieee_access_2023},
limiting fine-grained structured metadata for book organization and discovery.


Table~\ref{tab:literature_survey} summarizes representative book understanding
and genre analysis methods across different information sources and prediction
settings. Existing approaches exploit cover images \cite{judgingbookcover},
image-text representations \cite{exan_ijdar_2023,cover_iapr_2024},
title-image fusion \cite{cover_title_ICISGT_2019,benchmarking_cover}, blurbs
\cite{desc_bangla_2023_iccit,desc_spanish_ieee_access_2023}, reviews
\cite{book_review_Portuguese_Language2022}, and full-text chapters
\cite{book_chapters_2018_ICCL}. However, jointly exploiting blurbs and reviews
remains largely unexplored despite their complementary roles: blurbs provide
reliable but concise publisher-authored descriptions, whereas reviews offer
richer genre cues but may be subjective, sentiment-dominated, or irrelevant.
Moreover, most methods use flat single-label formulations with limited genre
coverage, with few considering multi-label prediction
\cite{desc_bangla_2023_iccit,desc_2017_ICMMI}. Thus, existing methods provide
limited support for integrating heterogeneous textual evidence within a
hierarchical multi-label organization.

\begin{table}[!t] 
    \centering
\caption{Representative studies on book organization}
    \vspace{-8pt}
    \begin{minipage}[t]{0.45\textwidth}
    \begin{adjustbox}{width=\textwidth}
        \begin{tabular}{l | l|  c| c| c| c} 
        \hline
          \textbf{Methods} & \textbf{Modality} & \textbf{Task} & \textbf{\# Genres} & \textbf{Multi-label} & \textbf{Hierarchy}  \\ \hline  

          \cite{genre_success_2020_saci} & Metadata & BSF & 609 & - & - \\ \hline
          
          \cite{judgingbookcover} & \multirow{2}{*}{Cover Image} &  BGP & 30 & \xmark & \xmark \\ 
          
          \cite{likability_2018_cemnlp} &  & BLE & 8 & - & - \\ \hline 
          

          \cite{cover_deep_multi_model_arxiv_2020} & \multirow{3}{*}{Cover (Image + Text)} & \multirow{3}{*}{BGP} & 30 & \xmark & \xmark \\ 
          \cite{cover_iapr_2024} & & & 30 & \xmark & \xmark \\
          \cite{exan_ijdar_2023} &  &  & 28 & \xmark & \xmark \\ \hline 

          \cite{benchmarking_cover} & \multirow{2}{*}{Cover Image + Title} & \multirow{2}{*}{BGP} & 30 & \xmark & \xmark \\ 
          \cite{cover_title_ICISGT_2019} &  &  & 6 & \xmark & \xmark \\ \hline    
     
          \cite{Title_ICCCS2019} & Title & BGP & 32 & \xmark & \xmark \\ \hline 
    
          \cite{desc_2017_ICMMI} & \multirow{3}{*}{Blurb} & \multirow{3}{*}{BGP} & 13 & \cmark & \xmark \\ 
          \cite{desc_bangla_2023_iccit} &  &  & 8 & \cmark & \xmark \\                             
          \cite{desc_spanish_ieee_access_2023} &  &  & 26 & \xmark & \xmark \\ \hline  

          \cite{book_chapters_2018_ICCL} & Book Text & BGP & 6 & \xmark & \xmark \\ \hline 
      
          \cite{book_review_Portuguese_Language2022} & \multirow{3}{*}{Reviews} & BGP & 24 & \xmark & \xmark \\ 
          \cite{review_2023_italian_jod} &  & WSA  & 6 & - & - \\ 
          \cite{rnn_review_2022_ijit} &  & BRec & 28 & - & - \\ \hline 
          
          \cite{best_seller_2017_icasbam} & Reviews + Rating & BSP & 24 & - & - \\ \hline
          
    
          \textbf{{\titleabbr}} (Ours) & {Reviews + Blurb} & {BGP} & {57} & \textbf{\cmark} & \textbf{\cmark} \\
          
           \hline
           \multicolumn{6}{r}{BRec: book recommendation, BSP: bestseller prediction, BLE: book likability estimation,} \\
           \multicolumn{6}{r}{BGP: book genre prediction, BSF: book success forecasting, and WSA: writing style analysis}\\
        \end{tabular}
    \end{adjustbox}
    \end{minipage}
    \label{tab:literature_survey}
\end{table}


Beyond book-specific methods, text classification has progressed from neural
models such as TextCNN \cite{textcnn_2014emnlp} to pretrained encoders such as
BERT \cite{bert} and graph-enhanced models such as BERTGCN
\cite{BertGCN_ACL_2021}. Recent multi-label text classification (MLTC) methods
model inter-label dependencies through semantic matching
\cite{DBC_MLTC_tkde_26}, graph-based correlations \cite{CorGCN_KDD25}, and
label-aware contrastive learning \cite{UCLAF_IJCAI_2025,BiG_ESA_2025}, while
hierarchical text classification (HTC) exploits label taxonomies through
hierarchy-aware graph reasoning and contrastive learning
\cite{HiAGM,hitin,hill,HiGATA_IJCNN_2025,LSIMCL_IPM_2026}. However, these
methods largely operate on unified textual representations without explicitly
modeling the distinct reliability and semantic roles of authoritative
descriptions and crowd-sourced interpretations. Moreover, conventional HTC
does not explicitly support progressive specialization from coarse book
categories to fine-grained multi-label characterization.

These limitations motivate {\titleabbr}, which combines blurb-guided semantic
refinement, semantic-role-separated dual-path graph convolutions, and structured
genre interaction modeling within a coarse-to-fine hierarchy. Unlike existing
book classification, MLTC, and HTC approaches, {\titleabbr} grounds noisy
crowd-sourced interpretations against authoritative descriptions, propagates
both sources independently, and models fine-grained inter-genre dependencies.
This produces reliable hierarchical book characterization and structured
semantic metadata for scalable book organization and learning-resource discovery.

\section{Solution Architecture}
\label{sec:method}

\subsection{Problem Formulation} \label{sec:problem}
Given a book corpus consisting of a set of $n$ books, $\mathcal{S} = \{\mathcal{S}_1, \mathcal{S}_2, \dots, \mathcal{S}_n\}$, where each book $\mathcal{S}_i$ is associated with a blurb $\mathcal{B}_i$, a set of $q$ user reviews $ \mathcal{R}_i = \{\mathcal{R}^i_1, \mathcal{R}^i_2, \ldots, \mathcal{R}^i_q\}$, and a two-level genre labels $\mathcal{Y}^i = \{\mathcal{Y}^i_1, \mathcal{Y}^i_2\}$, where $\mathcal{Y}^i_1 \in \{0, 1\}$ denotes the coarse label (\emph{fiction} or \emph{non-fiction}), and where $\mathcal{Y}^i_2$ denotes the fine-grained multi-label genres:

\vspace{-.2cm}

$$
\footnotesize
  	\mathcal{Y}^i_2 \in
        \begin{cases} 
            \{0,1\}^{k_1}; & \text{if } \mathcal{Y}^i_1 = 0 ~ \text{(\emph{fiction})}, \\ 
            \{0,1\}^{k_2}; & \text{if } \mathcal{Y}^i_1 = 1 ~ \text{(\emph{non-fiction})},
        \end{cases}
$$ 

\vspace{-.15cm}

\noindent
$k1$ and $k_2$ are fiction and non-fiction genre counts.
Let $\mathcal{B} = \{\mathcal{B}_1, \mathcal{B}_2, \dots, \mathcal{B}_n\}$ and $\mathcal{R} = \{\mathcal{R}_1, \mathcal{R}_2, \ldots, \mathcal{R}_n \}$.
The objective is to utilize book blurbs and user reviews to derive structured
genre metadata for learning-resource organization through two-stage inference: 
first, a Level-1 binary classification to distinguish fiction from non-fiction, and then a Level-2 multi-label classification of fine-grained genres conditioned on the Level-1 output.

\begin{figure}[!b]
    \includegraphics[width=\linewidth]{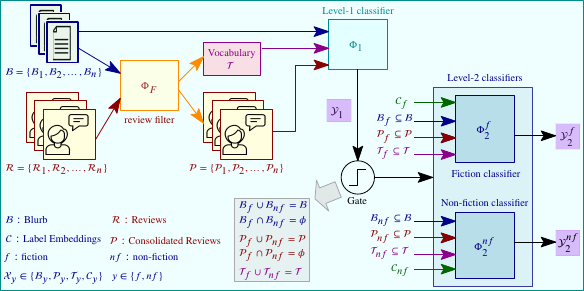}
    \caption{Overall framework of {\titleabbr}}
    \label{fig:overall_framework}
\end{figure}

Our framework, {\titleabbr} (Fig.~\ref{fig:overall_framework}), mitigates review noise through zero-shot semantic alignment that retains only blurb-consistent reviews, and formulates genre prediction as a two-stage hierarchical task using a dual-path graph learning architecture. The first stage performs fiction/non-fiction classification, which then activates the corresponding second-stage multi-label classifier for fine-grained genre prediction 
{(Appendix \ref{appendix:dataset})}.

\begin{figure}[!b]
    \includegraphics[width=\linewidth]{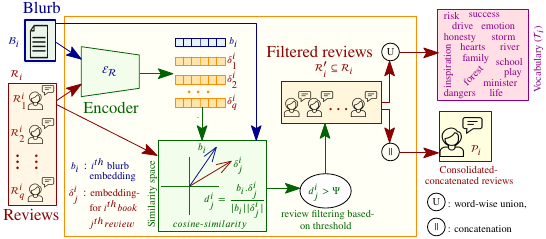}
    \caption{Review filtering and vocabulary generation}
    \label{fig:review_filtering_approach}
\end{figure}

\subsection{Blurb-Guided Semantic Refinement} 
\label{sec:review_filtering_approach}
Book blurbs and user reviews represent fundamentally different sources of literary semantics. Blurbs provide publisher-authored narrative intent, whereas reviews capture subjective reader interpretations and emotional reactions. Existing book genre classification methods predominantly rely on either blurbs or reviews independently, overlooking the complementary yet heterogeneous semantic information across these sources. Moreover, directly aggregating reviews often introduces semantic noise, sentiment bias, and genre-irrelevant interpretations during fine-grained genre inference. To address this challenge, {\titleabbr} introduces a \emph{blurb-conditioned semantic refinement} strategy that uses authoritative narrative descriptions to contextualize and refine crowdsourced review evidence before hierarchical genre reasoning. As illustrated in Fig.~\ref{fig:review_filtering_approach}, a BERT-based encoder $\mathcal{E}_{\mathcal{R}}$~\cite{bert} projects each blurb and review into a shared semantic space, producing embeddings $b_i$ and $\delta^i_j$, respectively. Standard preprocessing, including removal of emojis, hyperlinks, and repeated filler tokens, is applied before encoding. Semantic consistency between the blurb and each review is estimated using cosine similarity, and only reviews whose similarity score $d^i_j$ exceeds a dynamic threshold $\Psi$ are retained. The selected reviews are concatenated into a consolidated representation $\mathcal{P}_i$, preserving semantically grounded reader evidence while suppressing noisy and genre-irrelevant content. Since the refinement mechanism is nonparametric, it remains scalable, architecture-agnostic, and effective across varying review distributions. If a blurb is excessively short, all reviews are retained to avoid semantic under-specification. Finally, the refined blurbs and reviews are used to construct a shared vocabulary $\mathcal{T} = \{\mathcal{T}_1, \mathcal{T}_2, \ldots, \mathcal{T}_m\}$ for downstream genre-conditioned semantic representation learning.

\begin{figure*}[!t]
    \includegraphics[width=0.95\linewidth]{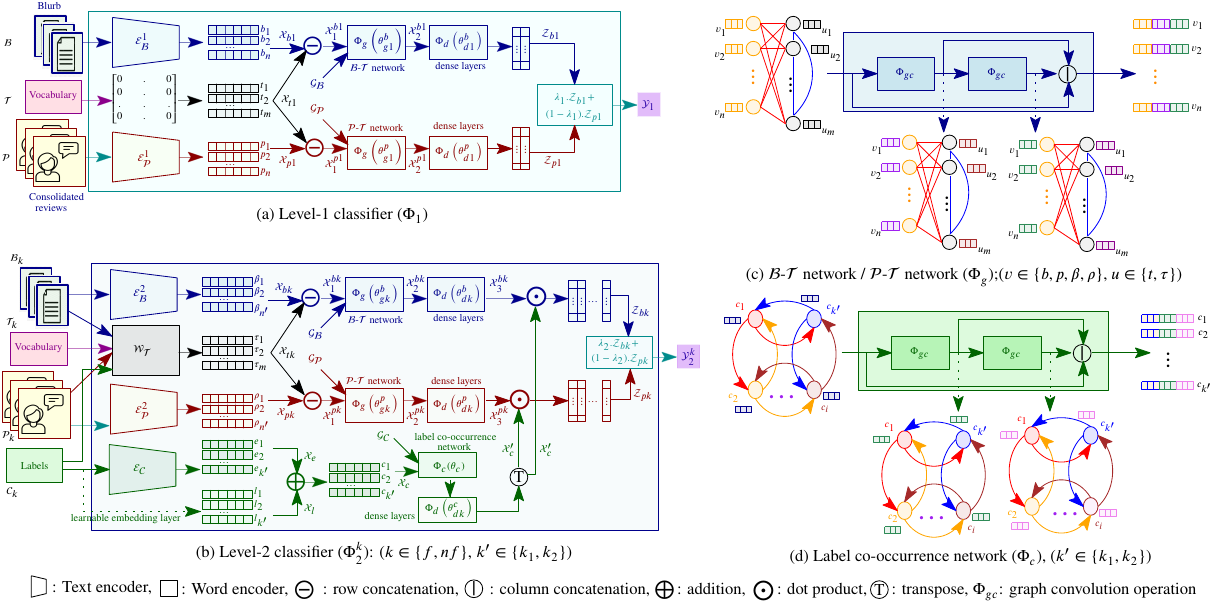}
    \caption{Detailed framework of {\titleabbr}. Best viewed in color.}
    \label{fig:detailed_arch}
\end{figure*}

\subsection{Semantic-Role-Separated Dual-Path Graph Reasoning}

In the second phase, {\titleabbr} performs hierarchical genre prediction using consolidated reviews and blurbs. Unlike conventional multimodal text classification frameworks that project all textual evidence into a unified semantic space, {\titleabbr} models blurbs and reviews as semantically heterogeneous evidence sources. Blurbs primarily encode publisher-authored narrative intent and thematic structure, whereas reviews capture subjective reader interpretations and experiential semantics. Joint propagation over a shared graph can therefore introduce semantic interference during hierarchical genre reasoning. 
To address this challenge, Level-1 classification utilizes the input triplet $(\mathcal{B}, \mathcal{P}, \mathcal{T})$ to train the model $\Phi_1$, where $\mathcal{T}$ denotes the shared token set derived from refined blurbs and reviews. As illustrated in Fig.~\ref{fig:detailed_arch}(a), $\Phi_1$ performs dual-path semantic-role-separated graph propagation through independent ${\mathcal{B}}\text{-}{\mathcal{T}}$ and ${\mathcal{P}}\text{-}{\mathcal{T}}$ reasoning branches, enabling separate modeling of narrative and interpretive semantics. The resulting representations are integrated through the weighting factor $\lambda_1$
to predict the coarse-grained label $\mathcal{Y}_1$.

\noindent
\textit{\textbf{Semantic-Role-Separated Graph Construction:}}
{\titleabbr} constructs independent blurb-token and review-token heterogeneous graphs to preserve semantic specialization during graph propagation.
Formally, we construct the blurb-token graph $\mathcal{G}_{\mathcal{B}}=(\mathbb{V}_{\mathcal{B}},\mathbb{E}_{\mathcal{B}})$ and the review-token graph $\mathcal{G}_{\mathcal{P}}=(\mathbb{V}_{\mathcal{P}},\mathbb{E}_{\mathcal{P}})$ over the refined textual inputs and the shared vocabulary $\mathcal{T}$.

The node sets $\mathbb{V}_{\mathcal{B}}=\{\mathcal{B},\mathcal{T}\}$ and $\mathbb{V}_{\mathcal{P}}=\{\mathcal{P},\mathcal{T}\}$ consist of document (i.e., blurb or review) and token nodes, while the edge sets encode document-token associations and token co-occurrence dependencies. Edge weights are computed using TF-IDF-based document-token importance and positive PMI-based token co-occurrence statistics: 
\begin{equation}
\scriptsize
\label{eq:edge_weights}
\mathcal{W}_{ij} =
\begin{cases} 
    f^t(\mathbb{V}_i, \mathbb{V}_j), & \text{if } \mathbb{V}_i {\in} \{\mathcal{B},\mathcal{P}\}, \mathbb{V}_j {\in} \mathcal{T} \\
    f^{p^{+}}(\mathbb{V}_i, \mathbb{V}_j)=\max{\left( \log \frac{p(\mathbb{V}_i, \mathbb{V}_j)}
    {p(\mathbb{V}_i)\, p(\mathbb{V}_j)}, 0 \right)}, 
    & \text{if } \mathbb{V}_i,\mathbb{V}_j {\in} \mathcal{T} \\ 
    0, & \text{otherwise.}
\end{cases}
\end{equation} 
Here, $f^t(\mathbb{V}_i, \mathbb{V}_j)$ represents the TF-IDF score \cite{textgcn_aaai_2019} and  
$f^{p^{+}}(\mathbb{V}_i, \mathbb{V}_j)$ computes the {positive point-wise mutual information (PMI)} \cite{bertgcn_ijcnlp_2021}.
TF-IDF captures modality-specific lexical importance, whereas positive PMI models statistical token co-occurrence structure within each propagation branch. The resulting adjacency matrices $\mathcal{A}_{\mathcal{B}}$ and $\mathcal{A}_{\mathcal{P}}$ therefore preserve narrative and interpretive semantic organization separately during graph reasoning.

\noindent
\textit{\textbf{Node Initialization for Level-1:}}  
To preserve modality-specific semantic specialization, {\titleabbr} employs independently fine-tuned encoder-based LMs $\mathcal{E}^1_{\mathcal{B}}$ and $\mathcal{E}^1_{\mathcal{P}}$ for blurbs and reviews, producing contextual representations: 
$\mathcal{X}_{b1}=[b_1,b_2,\dots,b_n]$, 
$\mathcal{X}_{p1}=[p_1,p_2,\dots,p_n]$, 
where $b_i$, $p_i \in \mathbb{R}^{h}$. The encoder parameters are subsequently frozen to stabilize downstream graph reasoning. 
Token nodes are initialized as: 
$\mathcal{X}_{t1}=[t_1,t_2,\dots,t_m]$, 
$t_i \in \{0\}^{j}$, allowing semantic interactions to emerge progressively through modality-specific graph propagation rather than static pretrained lexical priors.

\noindent 
\textit{\textbf{Semantic-Role-Separated Graph Propagation for Level-1:}} 
As illustrated in Fig.~\ref{fig:detailed_arch}(a), {\titleabbr} performs dual-path graph reasoning through modality-specific propagation branches for blurbs and reviews. Each branch consists of a graph propagation unit $\Phi_g$ followed by a lightweight feed-forward network $\Phi_d$. Rather than projecting all textual evidence into a unified representation space, the proposed architecture preserves narrative and interpretive semantics through independent graph propagation dynamics.

The graph propagation unit $\Phi_g$ captures structured semantic interactions by propagating contextual information over the heterogeneous text graphs. The graph propagation branches preserve complementary semantic interactions across blurbs and reviews during hierarchical reasoning.
%
Fig.~\ref{fig:detailed_arch}(c) illustrates the internal formulation of $\Phi_g$, consisting of two stacked graph convolution layers $\Phi_{gc}$:
%
%
\begin{equation}
    \footnotesize
    \label{eq:graph_convolution1}
    \begin{aligned}
          \Phi_{gc}(\hat{\mathcal{A}}, \mathcal{X}_1) &= \sigma (\hat{\mathcal{A}} {\cdot} \mathcal{X}_1{\cdot}\mathbb{W} + \theta), 
    \end{aligned}
\end{equation}
where {\scriptsize $\hat{\mathcal{A}} = \mathcal{D}^{-1/2} \cdot (\mathcal{A} + \mathbb{I}) \cdot \mathcal{D}^{-1/2}$} denotes the normalized adjacency matrix corresponding to $\mathcal{A}_{\mathcal{B}}$ or $\mathcal{A}_{\mathcal{P}}$, $\mathcal{D}$ is a diagonal matrix; {\scriptsize $\mathcal{D}_{ii} = \sum_{j} (\mathcal{A} + \mathbb{I})_{ij}$}. 
The identity matrix $\mathbb{I}$ is used to preserve the original features of each node. $\mathcal{X}_1$ is the node feature embedding matrix. 
Here, $\sigma$ is an activation function, while $\mathbb{W}$ and $\theta$ are learnable weight and bias parameters, respectively. 

To preserve both local and propagated semantics, {\titleabbr} concatenates the initial node features with the outputs of successive propagation layers:
\begin{align}
    \footnotesize
    \label{eq:graph_convolution11}
    \begin{split}
        \mathcal{X}^{k1}_{2} = T\left(\mathcal{X}^{k1}_{1} ~||~ \Phi_{gc}(.) ~||~ \Phi_{gc}(\Phi_{gc}(\cdot))\right);\quad k \in \{b, p\}
    \end{split}
\end{align}
%
%
where $T(\cdot)$ retains only document-node representations. 
By jointly preserving the initial semantic features and multi-hop propagated representations, the proposed aggregation strategy mitigates graph over-smoothing and over-squashing effects. Specifically, direct retention of the original node features prevents excessive homogenization of node representations across propagation layers, while concatenation of intermediate representations preserves complementary local and long-range semantic interactions without forcing them into a compressed latent space. This enables {\titleabbr} to maintain discriminative modality-specific semantics during hierarchical genre reasoning.
The resulting semantic representations are processed through $\Phi_d$ to generate $\mathcal{Z}_{b1}$ and $\mathcal{Z}_{p1}$ for the narrative and interpretive branches, respectively.
The final prediction $\mathcal{Y}_1$ is obtained through weighted evidence integration controlled by $\lambda_1$. If blurbs are insufficient, $\lambda_1=0$ suppresses narrative evidence; if reviews are insufficient, $\lambda_1=1$ suppresses interpretive evidence. Otherwise, $\lambda_1 \in (0,1)$ balances both sources based on their relative content completeness.

For optimization, the LM encoders are first fine-tuned and then frozen, while the remaining parameters are trained with Binary Cross-Entropy (BCE) loss~\cite{PyTorch}.
\subsection{Genre-Conditioned Hierarchical Refinement} 
\noindent
Level-2 performs domain-specialized multi-label genre reasoning through two mutually exclusive classifiers, $\Phi^f_2$ and $\Phi^{nf}_2$, corresponding to fiction and non-fiction domains, respectively. As illustrated in Fig.~\ref{fig:overall_framework}, the Level-1 prediction activates the appropriate Level-2 branch via a gating mechanism, enabling progressive semantic specialization from coarse-grained literary domains to fine-grained genre inference.

Given an input tuple $(\mathcal{B}_k,\mathcal{P}_k,\mathcal{T}_k,\mathcal{C}_k)$, where $k \in \{f,nf\}$, each classifier $\Phi^k_2$ performs multi-label genre prediction over semantically coherent domain-specific subspaces. Level-2 extends the Level-1 reasoning framework through genre-conditioned multi-label refinement.
In addition, Level-2 introduces structured genre interaction modeling through a label co-occurrence graph to capture fine-grained inter-genre dependencies.

To preserve domain-specific semantic specialization, separate encoder-based LMs $\mathcal{E}^k_{\mathcal{B}}$ and $\mathcal{E}^k_{\mathcal{P}}$, $k \in \{f, nf\}$, are independently fine-tuned for fiction and non-fiction multi-label objectives, producing contextual blurb and review representations:
$
\mathcal{X}_{bk}=[\beta_1,\dots,\beta_{n_k}] $, 
$
\mathcal{X}_{pk}=[\rho_1,\dots,\rho_{n_k}],
$
where $\beta_i, \rho_i \in \mathbb{R}^{h}$ and $n_k$ (number of books belonging to domain $k$) $< n$.

\noindent
\textit{\textbf{{Genre-Conditioned Token Semantics ${\mathcal{T}_k}$:}}} 
Unlike conventional lexical embeddings that remain genre-agnostic, {\titleabbr} constructs genre-conditioned token representations by encoding each word according to its normalized semantic distribution across genre categories. This enables the model to capture fine-grained associations between lexical usage patterns and hierarchical genre semantics.
To construct token embeddings $\mathcal{X}_{tk}=[\tau_1,\tau_2,\dots,\tau_m]$, where $\tau_i \in \mathbb{R}^{h}$, we aggregate the occurrences of each token $w_i \in \mathcal{T}_k$ across blurbs and reviews belonging to genre category $\mathcal{C}_j$: 
\begin{equation}
    \footnotesize
        \gamma_{ij}
        =
        \sum_{\mathcal{S}_k \in \mathcal{C}_j}\alpha_{ik}
        +
        \sum_{\mathcal{S}_k \in \mathcal{C}_j}\beta_{ik}~; \quad        
        \tilde{\gamma}_{ij}
        =
        \left|
        {(\gamma_{ij}-\bar{\gamma}_i)}/{\sigma_i}
        \right|,
\end{equation}
where $\alpha_{ik}$ and $\beta_{ik}$ denote the frequencies of $w_i$ in the blurb $\mathcal{B}_k$ and review $\mathcal{P}_k$, respectively, while $\bar{\gamma}_i$ and $\sigma_i$ represent the mean and standard deviation of the genre-wise frequency distribution of $w_i$. $\mathcal{S}_k \in \mathcal{C}_j$ denotes all books associated with genre $\mathcal{C}_j$. This shows that both blurb and review frequencies contribute to the genre-level representation of each word.
This formulation jointly captures narrative and interpretive genre associations while normalizing frequency variations across genre categories.

Let $\gamma_i = [\tilde{\gamma}_{i1}, \tilde{\gamma}_{i2}, \dots, \tilde{\gamma}_{ik'}]$ denote the normalized genre-distribution vector of token $w_i$.
Each token is then projected into the semantic embedding space using the genre embedding matrix $\mathcal{X}_e$, as described later:
$
\tau_i = \gamma_i {\cdot} \mathcal{X}_e.
$
Consequently, the resulting embedding $\tau_i$ explicitly encodes hierarchical genre affinity rather than relying solely on static lexical similarity, enabling more discriminative genre-aware semantic representations for fine-grained multi-label reasoning.

\noindent
\textit{\textbf{Structured Genre Interaction Modeling:}}
To capture fine-grained dependencies among genre categories, {\titleabbr} constructs a directed label interaction graph $\mathcal{G}_\mathcal{C}=(\mathbb{V}_\mathcal{C},\mathbb{E}_\mathcal{C})$, where nodes correspond to genre labels and edges encode conditional co-occurrence relationships derived from training data. Unlike independent label prediction, the proposed formulation explicitly models hierarchical semantic interactions among genres, enabling the framework to capture complementary and overlapping literary semantics. The edge weights are estimated using conditional probabilities $P(\mathcal{C}_i \mid \mathcal{C}_j)$ computed from label co-occurrence statistics, producing the correlation matrix $\mathcal{A}_\mathcal{C}$. To suppress noisy associations while emphasizing meaningful genre dependencies, $\mathcal{A}_\mathcal{C}$ is refined through threshold-based filtering using $\psi_1$ and $\psi_2$, preserving only semantically informative inter-label interactions. Specifically, connections with weights below a lower threshold $\psi_1$ are discarded, those above an upper threshold $\psi_2$ are reinforced by setting their values to 1, and intermediate values are retained as-is.

\noindent 
\textit{\textbf{{Genre Label Semantics for $\mathbb{V}_\mathcal{C}$:}}}
Genre labels are initialized using pretrained GloVe embeddings $\mathcal{X}_e=[e_1,e_2,\dots,e_{k'}]$, while an additional learnable embedding matrix $\mathcal{X}_l=[l_1,l_2,\dots,l_{k'}]$ enables task-specific semantic adaptation. The final genre representations are obtained as:
$
\mathcal{X}_c = \mathcal{X}_e + \mathcal{X}_l,
$
where $\mathcal{X}_c=[c_1,c_2,\dots,c_{k'}]$, where $k' \in \{k_1, k_2\}$. This hybrid formulation preserves pretrained semantic structure while allowing task-specific refinement of genre relationships during hierarchical multi-label reasoning.
\noindent
\textit{\textbf{Genre-Conditioned Hierarchical Reasoning:}} 
Fig.~\ref{fig:detailed_arch}(b) illustrates the Level-2 architecture. While the $\mathcal{B}\text{-}\mathcal{T}$ and $\mathcal{P}\text{-}\mathcal{T}$ branches follow the semantic-role-separated propagation framework introduced in Level-1, Level-2 further incorporates structured genre interaction reasoning through the label graph network $\Phi_c$ (Fig.~\ref{fig:detailed_arch}(d)). By propagating information over the genre interaction graph $\mathcal{G}_\mathcal{C}$ using genre-aware label representations $\mathcal{X}_c$, $\Phi_c$ learns refined genre semantics $\mathcal{X}'_c$ that capture fine-grained inter-genre dependencies essential for multi-label prediction. 
The refined genre representations are subsequently aligned with the narrative and interpretive semantic branches through dot-product interactions:
\begin{equation}
    \label{eq:in_level_2}
    \footnotesize
        \mathcal{Z}_{bk}
        =
        \mathcal{X}^{bk}_{3}{\cdot}\mathcal{X}'_c~;~
        \mathcal{Z}_{pk}
        =
        \mathcal{X}^{pk}_{3}{\cdot}\mathcal{X}'_c~;~
        \mathcal{Y}^k_2
        =
        \lambda_2\mathcal{Z}_{bk}
        +
        (1-\lambda_2)\mathcal{Z}_{pk}.
\end{equation}
The outputs of the narrative and interpretive propagation branches after graph reasoning and dense transformation are denoted as $\mathcal{X}^{bk}_{3}$ and $\mathcal{X}^{pk}_{3}$, respectively.
The weighting parameter $\lambda_2$ performs reliability-aware integration between authoritative narrative evidence and crowdsourced interpretive evidence based on source completeness. Final genre predictions are obtained through sigmoid activation followed by threshold-based label assignment.
The Level-2 classifiers are optimized using BCE with logits loss. To prevent hierarchical error propagation, the fiction and non-fiction losses are gated using the ground-truth Level-1 label, ensuring that gradient updates are restricted to the semantically relevant branch during training.

\section{Experiments}

\subsection{Dataset and Experimental Setup}
\label{subsec:datasets}

To the best of our knowledge, no public dataset currently supports hierarchical
book genre classification with paired blurbs and user reviews. We therefore
construct a two-level hierarchical genre dataset using BookCrossing
\cite{bookcrossing2005}, augmented with book blurbs and user reviews crawled
from Goodreads. Level-1 comprises \emph{fiction} and \emph{non-fiction}, while
Level-2 contains fine-grained sub-genres. With assistance from linguistic
experts, semantically overlapping genres are consolidated to reduce redundancy.
Dataset statistics are reported in Table~\ref{tab:dataset}, with the detailed
taxonomy and genre distributions provided in Appendix~\ref{appendix:dataset}.

Experiments were conducted in PyTorch 2.5.1 using a 7:1:2
train/validation/test split. To mitigate Level-2 class imbalance, we applied
Gemini-based data augmentation \cite{gemini2_5_flash2025}. Detailed
augmentation statistics and model configurations are provided in
Appendices~\ref{appendix:dataset} and
\ref{appendix:implementation_and_hyperparameter_details}, respectively.
All results are reported on the test set. Level-1 is evaluated using F1-score
($\mathcal{F}$) and Accuracy ($\mathcal{A}_c$), while Level-2 uses micro/macro
F1 (${\mathcal{F}}\mu$, ${\mathcal{F}}m$), Balanced Accuracy
(${\mathcal{BA}}\mu$, ${\mathcal{BA}}_m$), and Hamming Loss
($\mathcal{HL}$) for imbalance-aware multi-label evaluation.

\begin{table}[!hbt]
    \centering
    \caption{Dataset overview}
    \begin{adjustbox}{width=0.48\textwidth}
        \begin{tabular}{l | c|c|c | c|c|c} \hline
           \textbf{Genre} & \textbf{No. of} & \textbf{No. of} & \textbf{No. of} & \textbf{Avg. reviews} & \textbf{Avg. words} & \textbf{Avg. words} \\  
           \textbf{categories} & \textbf{books} & \textbf{genres} & \textbf{reviews} & \textbf{per book} & \textbf{per blurb} & \textbf{per review} \\ \hline
           Fiction & 5612 & 28 & 54651 & 9.74 & 100.35 & 183.13 \\
           Non-Fiction & 3763 & 29 & 35848 & 9.53 & 112.89 & 117.79 \\  \hline 
        \end{tabular}
    \end{adjustbox} 
    \label{tab:dataset}
\end{table}

\newcommand{\tableheader}{%
  \cline{2-12}
  \multirow{2}{*}{\textbf{}} & \multirow{2}{*}{\textbf{Models}}
    & \multicolumn{2}{c|}{\textbf{Level-1}}
    & \multicolumn{4}{c|}{\textbf{Level-2: Fiction (F)}}
    & \multicolumn{4}{c}{\textbf{Level-2: Non-Fiction (NF)}} \\ \cline{3-12}
    & & \(\mathcal{F}\uparrow\) & \(\mathcal{A}_c\uparrow\)
    & \(\mathcal{F}_\mu\uparrow\) & \(\mathcal{BA}_\mu\uparrow\)
    & \(\mathcal{F}_m\uparrow\)  & \(\mathcal{BA}_m\uparrow\)
    & \(\mathcal{F}_\mu\uparrow\) & \(\mathcal{BA}_\mu\uparrow\)
    & \(\mathcal{F}_m\uparrow\)  & \(\mathcal{BA}_m\uparrow\) \\ \hline
}

\begin{table*}[!t]
\centering
\scriptsize
\caption{Performance of {\titleabbr} with traditional and LM-based methods}
\centering
\begin{adjustbox}{width=0.70\textwidth}
\begin{tabular}{l | l | cc | cccc | cccc}
  \tableheader

   \multirow{1}{*}{SOTA} & Scofield et al. \cite{book_review_Portuguese_Language2022} & -- & -- & 11.99 & 53.19 & 5.97  & 51.57 & 7.80  & 52.03 & 10.12 & 52.77 \\
    (only & N.-Flores et al. \cite{desc_spanish_ieee_access_2023} & -- & -- & 29.85 & 59.09 & 14.18 & 54.56 & 27.86 & 58.28 & 24.30 & 57.69 \\
    blurb) & Ullah et al. \cite{desc_bangla_2023_iccit} & --& --& \underline{39.62} & \underline{63.28} & \underline{18.35} & \underline{56.31}
                               & \underline{39.84} & \underline{63.18} & \underline{32.55} & \underline{61.30} \\ 
\cline{2-12}

     & \cellcolor{black!20} Gain (\%)
                        & \cellcolor{black!20}-- &    \cellcolor{black!20}--
                           & \cellcolor{black!20}106.69 & \cellcolor{black!20}46.65
                           & \cellcolor{black!20}320.93 & \cellcolor{black!20}64.25
                           & \cellcolor{black!20}105.47 & \cellcolor{black!20}48.78
                           & \cellcolor{black!20}141.38 & \cellcolor{black!20}52.09 \\ 
\hline 

    SOTA & Ullah et al.  \cite{desc_bangla_2023_iccit}   & -- & -- & 43.65 & 65.03 & 21.40 & 57.33 & 47.65 & 67.02 & 39.69 & 64.54 \\
   (only & Scofield et al. \cite{book_review_Portuguese_Language2022} & -- & -- & 50.89 & 67.29 & \underline{41.71} & \underline{64.04}
                              & 40.29 & 62.70 & 33.64 & 60.68 \\
   review) & N.-Flores et al. \cite{desc_spanish_ieee_access_2023}& -- & -- & \underline{53.05} & \underline{69.66} & 35.25 & 63.99
                              & \underline{60.14} & \underline{75.05} & \underline{56.65} & \underline{74.19} \\ \cline{2-12}
    & \cellcolor{black!20} Gain (\%)
                       & \cellcolor{black!20}--  & \cellcolor{black!20}-- & \cellcolor{black!20}54.36 & \cellcolor{black!20}33.22
                           & \cellcolor{black!20}85.18 & \cellcolor{black!20}44.43
                           & \cellcolor{black!20}36.12 & \cellcolor{black!20}25.25
                           & \cellcolor{black!20}38.69 & \cellcolor{black!20}25.66 \\ \hline

  \multirow{4}{*}{MLTC}
    & DBC  \cite{DBC_MLTC_tkde_26}  & -- & -- & 20.59 & 55.49 & 18.45 & 51.61 & 15.42 & 53.55 & 16.54 & 53.67 \\
    & UCLAF \cite{UCLAF_IJCAI_2025} & -- & -- & 80.20 & \underline{90.82} & 75.68 & \underline{89.46} & \underline{79.50} & \underline{87.92} & \underline{76.92} & \underline{89.36} \\
    & BiG \cite{BiG_ESA_2025}  & -- & -- & \underline{80.52} & 88.95 & \underline{75.81} & 87.78
                     & 78.15 & 85.11 & 75.77 & 83.86 \\ \cline{2-12}
    & \cellcolor{black!20} Gain (\%)
     & \cellcolor{black!20}-- & 
     \cellcolor{black!20}-- & 
     \cellcolor{black!20}1.70 & 
     \cellcolor{black!20} 2.18
     & \cellcolor{black!20} 1.89 & 
     \cellcolor{black!20} 3.39
     & \cellcolor{black!20} 2.97 & 
     \cellcolor{black!20} 6.92
     & \cellcolor{black!20} 2.15 & \cellcolor{black!20} 4.33 \\ \hline

  \multirow{4}{*}{HTC}
    & HiTIN \cite{hitin} & 96.85 & 96.21 & 73.84 & 86.04 & 62.90 & 80.88 & 74.40 & 86.88 & 68.90 & 84.47 \\
    & HILL \cite{hill}  & 96.94 & \textbf{97.06} & 75.90 & 83.70 & 76.23 & 84.79 & 76.19 & 83.53 & 76.73 & 85.06 \\
    & L-SIMCL \cite{LSIMCL_IPM_2026} &  \underline{97.42} &  96.88 & \underline{78.12} & \underline{89.89} & \underline{76.86} & \underline{90.89}
      & \underline{79.74} & \underline{90.33} & \underline{77.93} & \underline{91.13} \\ \cline{2-12}
      & \cellcolor{black!20} Gain (\%)
      & \cellcolor{black!20} 0.08
      & \cellcolor{black!20} -0.15 
      & \cellcolor{black!20}4.83 
      & \cellcolor{black!20}3.24
      & \cellcolor{black!20}0.49 
      & \cellcolor{black!20}1.76
      & \cellcolor{black!20}2.66
      & \cellcolor{black!20}4.06
      & \cellcolor{black!20}0.82
      & \cellcolor{black!20}2.30 \\ \hline

   & TextGCN \cite{textgcn_aaai_2019} & -- & -- & 26.73 & 64.12 & 21.18 & 54.77 & 28.41 & 67.02 & 33.06 & 64.33  \\ 
   
   & BertGCN \cite{bertgcn_ijcnlp_2021} & -- & -- & 50.82 & \underline{86.98} & 58.20 & 85.08 & 54.39 & 69.32 & 53.92 & 71.39  \\ 
   
   \multirow{1}{*}{Graph-} & CorGCN \cite{CorGCN_KDD25} & -- & -- & 67.70 & 76.54 & 50.30 & 71.32 & \underline{75.76} & \underline{83.09} & 73.70 & 83.26 \\
    
    based & HiAGM \cite{HiAGM}  & 95.34& 94.39 & 64.13 & {80.47} & 47.95 & 73.31 & 64.04 & 80.13 & 57.82 & 78.63 \\
     & HiGATA \cite{HiGATA_IJCNN_2025} & \underline{97.33} & \underline{96.82} & \underline{70.68} & 79.51 & \underline{74.97} & \underline{86.15} & 67.51 & 81.58 & \underline{77.11} & \underline{87.07} \\
     \cline{2-12}
    & \cellcolor{black!20} Gain (\%)
     & \cellcolor{black!20} 0.17 & \cellcolor{black!20} 0.09 & \cellcolor{black!20}15.86 & \cellcolor{black!20}6.69
         & \cellcolor{black!20}3.03 & \cellcolor{black!20}7.36
         & \cellcolor{black!20}8.05 & \cellcolor{black!20}13.13
         & \cellcolor{black!20}1.89 & \cellcolor{black!20}7.07 \\ \hline

  
    & ALBERT  \cite{ALBERT}       & 94.93 & 93.99 & 58.00 & 78.60 & 47.25 & 72.95 & 63.75 & 84.86 & 61.49 & 81.80 \\
    & XLNET  \cite{XLNet}        & 95.62 & 94.84 & 63.46 & 81.24 & 52.26 & 74.40 & 69.68 & 86.05 & 66.11 & 83.04 \\
    & BERT \cite{bert} & 96.05& 95.27  & 65.23 & 82.02 & 57.43 & 76.92 & 72.31 & \underline{88.16} & 67.53 & 83.82 \\
    {LM-} & DistilBERT  \cite{distilBERT}   & \underline{96.13} & \underline{95.37} & {67.39} & \underline{83.35} & {55.78} & {77.45}
                           & \underline{73.25} & 87.00 & \underline{68.77} & \underline{84.23} \\
    based & M-BERT  \cite{modernBERT}   & 95.60 & 94.36 & 59.94 & 79.02 & 50.94 & 73.28 & 69.51 & 86.62 & 65.11 & 82.98 \\
    & M-BERT-L  \cite{modernBERT}  & 93.98 &	94.87 &	65.24 &	81.37 &	59.38 &	19.45 & 70.59 &	86.92 &	66.33 &	82.64  \\
    & RoBERTa   \cite{RoBERTa}  & 95.16 & 94.15 & 65.88 & 80.89 & 51.49 & 74.76 & 72.07 & 87.97 & 66.46 & 83.77 \\
    & RoBERTa-L  \cite{RoBERTa} & 94.36 &	95.13 &	\underline{70.34} &	83.10 &	\underline{62.71} &	\underline{83.50} & 70.10 &	86.32 &	65.19 &	82.66  \\ \cline{2-12}
    & \cellcolor{black!20} Gain (\%)
     & \cellcolor{black!20} 3.33 & \cellcolor{black!20} 1.87 & \cellcolor{black!20} 16.42 & \cellcolor{black!20} 11.34
     & \cellcolor{black!20} 23.17 & \cellcolor{black!20} 10.77
    & \cellcolor{black!20} 11.75 & \cellcolor{black!20} 6.62
    & \cellcolor{black!20} 14.25 & \cellcolor{black!20} 10.69 \\ \hline

      \multicolumn{2}{c|}{{\titleabbr}}
        & \textbf{97.50} & \textbf{96.51} & \textbf{81.89} & \textbf{92.80} & \textbf{77.24} & \textbf{92.49}
            & \textbf{81.86} & \textbf{94.00} & \textbf{78.57} & \textbf{93.23} \\ 
            


        \multicolumn{2}{c|}{$\pm$ stdev}
& $\pm 0.06$ & $\pm 0.05$ & $\pm 0.49$ & $\pm 0.18$ & $\pm 0.68$ & $\pm 0.16$ & $\pm 0.41$ & $\pm 0.19$ & $\pm 0.64$ & $\pm 0.16$ \\ \hline

  \multicolumn{12}{r}{Category-wise, the second-best result is \underline{underlined}}
\end{tabular}
\end{adjustbox}
%
\label{tab:baseline_results_main}
\end{table*}

\newcommand{\llmtableheader}{%
  \cline{3-11}
  \multicolumn{1}{c}{} & \multirow{2}{*}{\textbf{}} & \multirow{2}{*}{\textbf{Models}}
    & \multicolumn{4}{c|}{\textbf{Level-2: Fiction (F)}}
    & \multicolumn{4}{c}{\textbf{Level-2: Non-Fiction (NF)}} \\ \cline{4-11}
    \multicolumn{1}{c}{}& & & \(\mathcal{F}_\mu\uparrow\) & \(\mathcal{BA}_\mu\uparrow\)
    & \(\mathcal{F}_m\uparrow\)  & \(\mathcal{BA}_m\uparrow\)
    & \(\mathcal{F}_\mu\uparrow\) & \(\mathcal{BA}_\mu\uparrow\)
    & \(\mathcal{F}_m\uparrow\)  & \(\mathcal{BA}_m\uparrow\) \\ \hline
}

\begin{table*}[!t]
\caption{Comparison of {\titleabbr} with LLMs}
\label{tab:sota_llm}
    \centering
    \begin{adjustbox}{width=0.70\textwidth}
    \begin{tabular}{l | l | l | cccc | cccc}
      \llmtableheader
    
      
        \multirow{12}{*}{\shortstack[l]{Open-\\source}} & \multirow{4}{*}{zero-shot} & Mistral-Small-3.2-24B-Instruct \cite{mistral7b2023}& 45.29 & 79.89 & 40.40 & \underline{77.90} & 47.96 & 75.70 & 29.62 & 65.58 \\
       &  & Qwen2.5-72B-Instruct  \cite{qwen3} & 52.86 & 79.36 & 45.73 & 77.24 & \underline{54.33} & 74.49 & \underline{33.97} & 66.52 \\
        & & Llama-3.3-70B-Instruct \cite{llama} & \underline{56.06} & \underline{82.24} & \underline{46.97} & 77.83 & 53.37 & \underline{77.50} & 32.82 & \underline{67.54} \\ \cline{3-11}
        & & \cellcolor{black!20} Gain (\%)
          & \cellcolor{black!20}46.08 & \cellcolor{black!20}12.84
              & \cellcolor{black!20}64.45 & \cellcolor{black!20}18.73
              & \cellcolor{black!20}50.67 & \cellcolor{black!20}21.29
              & \cellcolor{black!20}131.29 & \cellcolor{black!20}38.04 \\ \cline{2-11}
    
      & \multirow{4}{*}{\shortstack[l]{few-shot}}
        & Llama-3.3-70B-Instruct     \cite{llama}       & 76.60 & 85.86 & 72.04 & 85.13 & 73.03 & 82.50 & 54.88 & 74.98 \\
        & & Qwen2.5-72B-Instruct    \cite{qwen3}        & 77.50 & 86.14 & 73.23 & 86.39 & 77.11 & 85.49 & \underline{63.74} & 80.74 \\
        & & Mistral-Small-3.2-24B-Instruct  \cite{mistral7b2023}  & \underline{81.25} & \underline{90.08} & \underline{75.66} & \underline{88.65}
        & \underline{79.71} & \underline{88.43} & 62.53 & \underline{81.34} \\ \cline{3-11}
        & & \cellcolor{black!20} Gain (\%)
          & \cellcolor{black!20}0.79 & \cellcolor{black!20}3.02
      & \cellcolor{black!20}2.09 & \cellcolor{black!20}4.33
      & \cellcolor{black!20}2.70 & \cellcolor{black!20}6.30
      & \cellcolor{black!20}23.27 & \cellcolor{black!20}14.62 \\ \cline{2-11}
    
      & \multirow{4}{*}{\shortstack[l]{fine-tuned}}
        & Qwen3-4B  \cite{qwen3}  & 74.71 & 82.23 & 54.42 & 72.86 & 74.19 & 83.30 & 66.60 & 78.78 \\
        & & Gemma2-2B \cite{gemma2024}  & 75.45 & 83.70 & 58.41 & 76.22 & 75.24 & 83.48 & 70.12 & 80.51 \\
        & & Mistral-7B \cite{mistral7b2023} & \underline{81.20} & \underline{87.90} & \underline{69.51} & \underline{81.43}
        & \underline{78.79} & \underline{86.68} & \underline{73.27} & \underline{83.86} \\ \cline{3-11}
        & & \cellcolor{black!20} Gain (\%)
      & \cellcolor{black!20}0.85 & \cellcolor{black!20}5.57
      & \cellcolor{black!20}11.12 & \cellcolor{black!20}13.58
      & \cellcolor{black!20}3.90 & \cellcolor{black!20}8.44
      & \cellcolor{black!20}7.23 & \cellcolor{black!20}11.17 \\ \hline
    
      \multirow{8}{*}{\shortstack[l]{Closed-\\source}}
        & \multirow{4}{*}{\shortstack[l]{zero-shot}} & Gemini-2.5-Flash \cite{gemini2_5_flash2025} & 59.11 & \underline{84.41} & 51.79 & 82.85 & 62.57 & \underline{81.73} & 39.03 & \underline{72.26} \\
        &  & DeepSeek-v3-chat \cite{deepseek-v3} & 62.97 & 82.97 & 57.25 & \underline{83.16} & 64.75 & 80.03 & \underline{41.70} & 70.47 \\
        & & GPT-4.1-mini \cite{openai2025gpt4.1mini}    & \underline{64.08} & 81.89 & \underline{58.70} & 82.36
                                  & \underline{65.01} & 80.38 & {41.10} & 70.15 \\ \cline{3-11}
        & & \cellcolor{black!20} Gain (\%)
          & \cellcolor{black!20}27.79 & \cellcolor{black!20}9.94
          & \cellcolor{black!20}31.58 & \cellcolor{black!20}11.22
          & \cellcolor{black!20}25.92 & \cellcolor{black!20}15.01
          & \cellcolor{black!20}88.42 & \cellcolor{black!20}29.02 \\  \cline{2-11}
    
      & \multirow{4}{*}{\shortstack[l]{few-shot}}
        & Gemini-2.5-Flash \cite{gemini2_5_flash2025} & 78.54 & \textbf{94.60} & 72.93 & \textbf{94.75} & \underline{81.83} & \underline{93.78} & 67.97 & \underline{92.49} \\
        & & DeepSeek-v3-chat \cite{deepseek-v3} & \textbf{83.27} & 93.32 & \textbf{77.41} & 92.05 & 81.82 & 90.46 & \underline{70.79} & 86.48 \\
        & & GPT-4.1-mini \cite{openai2025gpt4.1mini}  & 76.97 & 86.09 & 69.68 & 83.81 & 75.07 & 83.79 & 60.04 & 78.47 \\ \cline{3-11}
        & & \cellcolor{black!20} Gain (\%)
          & \cellcolor{black!20}-1.66 & \cellcolor{black!20}-1.90
              & \cellcolor{black!20}-0.22 & \cellcolor{black!20}-2.39
              & \cellcolor{black!20}0.04 & \cellcolor{black!20}0.23
              & \cellcolor{black!20}10.99 & \cellcolor{black!20}0.80 \\ \hline
    


 \multicolumn{3}{c|}{{\titleabbr}}
        & \textbf{81.89} & \textbf{92.80} & \textbf{77.24} & \textbf{92.49} & \textbf{81.86} & \textbf{94.00} & \textbf{78.57} & \textbf{93.23} \\ 

        \multicolumn{3}{c|}{$\pm$ stdev}
        & $\pm0.49$ & $\pm0.18$ & $\pm 0.68$ & $\pm 0.16$
        & $\pm 0.41$ & $\pm 0.19$ & $\pm 0.64$ & $\pm 0.16$ \\ \hline

\multicolumn{11}{r}{Category-wise, the second-best result is \underline{underlined}}
    \end{tabular}
    \end{adjustbox}
\end{table*}

\subsection{Results \& Comparative Study}
To benchmark {\titleabbr}, we evaluate it against a diverse set of baselines, as shown in 
Tables~\ref{tab:baseline_results_main}-\ref{tab:sota_llm}.

\emph{(i) Past Book Genre Classification Methods:}
We include three representative book genre prediction approaches based on reviews and book descriptions. Scofield et al. \cite{book_review_Portuguese_Language2022} perform genre prediction by aggregating online user reviews and evaluating multiple conventional machine learning classifiers for flat genre classification. N.-Flores et al.~\cite{desc_spanish_ieee_access_2023} leverage book descriptions in Spanish and formulate genre categorization as a flat multi-class problem over 26 categories derived from a Latin American-customized Thema taxonomy, employing back-translation for data augmentation and evaluating transformer-based models including BERT and RoBERTa. Finally, Ullah et al. \cite{desc_bangla_2023_iccit} introduce a multi-label Bangla book classification framework using contextual book descriptions with deep neural and pretrained language models. For fair comparison, all methods were adapted to a unified 57-class setting (28 Fiction, 29 Non-Fiction) and evaluated separately using blurbs or reviews. As shown in the first segment of Table~\ref{tab:baseline_results_main}, prior unimodal methods generally struggle on macro-level performance. Review-based models consistently outperformed blurb-only counterparts at Level-2, indicating the richer information contained in user reviews. In contrast, by jointly leveraging blurbs and filtered reviews, {\titleabbr} achieved the best performance across all metrics. These results highlight the effectiveness of review filtering, multimodal fusion, and hierarchical label modeling for fine-grained genre classification.

\emph{(ii) Multi-label Text Classification (MLTC) Methods:} 
We further compare {\titleabbr} against three recent flat MLTC approaches, as shown in Table~\ref{tab:baseline_results_main}. DBC \cite{DBC_MLTC_tkde_26} performs distance-based semantic matching with adaptive threshold optimization. UCLAF \cite{UCLAF_IJCAI_2025} combines label-aware representation learning with contrastive optimization, while BiG \cite{BiG_ESA_2025} employs bidirectional document-label contrastive learning to model partial label overlap. 
All MLTC methods were adapted to a unified 57-class setting (28 Fiction and 29 Non-Fiction genres). For multimodal evaluation, blurbs and reviews were combined using either feature addition or concatenation, and the better-performing configuration was reported for each method.
Among them, UCLAF and BiG achieved competitive baseline performance. However, these methods operate in a flat prediction space and do not explicitly capture hierarchical semantic dependencies between coarse and fine-grained genres. 
In contrast, {\titleabbr} leverages hierarchical reasoning and graph-based multimodal learning, consistently outperforming all MLTC baselines across evaluation settings.

\emph{(iii) Hierarchical Text Classification (HTC) Methods:} 
We further extend the comparison to three recent hierarchy-aware HTC frameworks. HiTIN \cite{hitin} models hierarchical syntactic dependencies using coding-tree representations, while HILL \cite{hill} introduces hierarchy-aware information-lossless contrastive learning for improved hierarchical representations. L-SIMCL \cite{LSIMCL_IPM_2026} further combines label-guided semantic interaction with hierarchy-aware contrastive optimization to learn discriminative hierarchical features. Although these methods are designed for deeper taxonomies, all approaches are evaluated under the same two-level hierarchy setting of our dataset for fair comparison. 
For multimodal evaluation, blurbs and reviews were integrated using both feature addition and concatenation strategies, and the best-performing setting was reported for each HTC method.
As shown in Table~\ref{tab:baseline_results_main}, among them, L-SIMCL achieves the strongest baseline performance, demonstrating the effectiveness of hierarchy-aware contrastive learning for HTC. Furthermore, these methods primarily focus on modeling label hierarchies within a unified prediction space, without explicitly separating high-level semantic domains before fine-grained prediction. Consequently, they remain vulnerable to semantic interference across heterogeneous parent categories and struggle to capture the distinct semantic characteristics of fiction and non-fiction genres. In contrast, {\titleabbr} performs coarse-to-fine genre prediction through hierarchical reasoning and multimodal graph learning, consistently outperforming all HTC baselines across both \emph{fiction} and \emph{non-fiction} settings.

\emph{(iv) Graph-based Methods for Text Classification:} 
We include five representative graph-based frameworks covering both flat multi-label and hierarchical text classification settings. TextGCN \cite{textgcn_aaai_2019} is initialized with one-hot representations for words and documents, then jointly learns embeddings for both through graph convolution, while BertGCN \cite{BertGCN_ACL_2021} constructs a heterogeneous graph over the dataset and represents document nodes using BERT representations. Moreover, CorGCN~\cite{CorGCN_KDD25} employs correlation-aware graph decomposition and correlation-enhanced graph convolution to model inter-label dependencies for multi-label {node} classification on graph-structured data. For hierarchical modeling in graph-based learning settings, HiAGM \cite{HiAGM} captures hierarchical label dependencies through directed graph encoders with hierarchy-aware feature propagation, and HiGATA \cite{HiGATA_IJCNN_2025} further incorporates persistent homology-based graph attention learning to model topological characteristics of hierarchical label spaces. Similar to the earlier methods, graph-based approaches also integrated blurbs and reviews using both feature addition and concatenation strategies, and the best-performing configuration was reported for each method.
Although HiGATA and CorGCN achieved competitive second-best results (refer to Table~\ref{tab:baseline_results_main}), existing graph-based baselines primarily focus on flat multilabel classification or hierarchy-centric label modeling, without explicitly distinguishing the semantic roles of authoritative blurbs and subjective reviews, or addressing the semantic noise and genre-irrelevant interpretations commonly present in crowd-sourced reviews. {\titleabbr} addresses this gap by jointly encoding hierarchical semantics, cross-modal interactions, and curated review information, achieving consistently stronger performance in both \emph{fiction} and \emph{non-fiction} categorization.

\emph{(v) Language Model (LM)-based Methods:}
We employ six prominent LM-based baselines, including ALBERT \cite{ALBERT}, BERT \cite{bert}, DistilBERT \cite{distilBERT}, XLNet \cite{XLNet}, ModernBERT \cite{modernBERT}, and RoBERTa \cite{RoBERTa}, where the latter two are evaluated using both base and large variants. Details of the adopted encoders are provided in Sec.~\ref{appendix:encoder_selection}. For fair comparison, all models follow the same hierarchical pipeline, where the LM encoder is coupled with a two-layer feed-forward classifier at both Level-1 and Level-2. To integrate blurb and review representations, we evaluate multiple fusion strategies, including feature addition, concatenation, self-attention, and cross-attention, and report the best empirical results in Table~\ref{tab:baseline_results_main} of the main paper. {\titleabbr} consistently outperformed all LM-based methods, achieving notable gains over strong baselines such as BERT and DistilBERT. These improvements highlight the benefit of graph-based label dependency learning, review filtering, and contextual semantic modeling.

\begin{table*}[!t]
\centering
\caption{Modalities and module ablation study}
\begin{adjustbox}{width=0.70\textwidth}
    \centering
    \begin{tabular}{l|c|c|c|c|c|c|c|c|c}
    \hline
     
    \multirow{2}{*}{\textbf{Models}} & \multirow{2}{*}{\shortstack[l]{\textbf{Modality}}} & \multicolumn{2}{c|}{\textbf{Level-1}} & \multicolumn{3}{c|}{\textbf{Level-2: Fiction}}& \multicolumn{3}{c}{\textbf{Level-2: Non-Fiction}} \\ \cline{3-10} 
    
     & & \multicolumn{1}{c|} {\textbf{${\mathcal{F}}\uparrow$}} & \textbf{${\mathcal{A}c}\uparrow$} & \multicolumn{1}{c|}{\textbf{${\mathcal{F}}_{\mu}\uparrow$}} & \multicolumn{1}{c|}{\textbf{${\mathcal{F}}_{m}\uparrow$}} &  \multicolumn{1}{c|}{\textbf{${\mathcal{HL}}\downarrow$}} & \multicolumn{1}{c|}{\textbf{${\mathcal{F}}_{\mu}\uparrow$}} & \multicolumn{1}{c|}{\textbf{${\mathcal{F}}_{m}\uparrow$}}  & \textbf{${\mathcal{HL}}\downarrow$} \\ \hline 
    
    {(a)} {\titleabbr} (1, 1) & \textbf{b}  & 94.71 & 95.80 & 66.87 & 55.07 & 0.0886 & 72.64 & 65.30 & 0.0676 \\ 
    
    {(b)} {\titleabbr} (0, 0) & \textbf{r} & 87.80 & 90.69 & 74.90 & 63.48 & 0.0641 & 75.23 &  70.65 &  0.0598 \\ 
    
    {(c)} {\titleabbr} (0.5, 0.5) & \textbf{b + r} & 96.71 & 97.34 & 78.91 & 70.01 & 0.0517 & 78.17 & 73.36 &  0.0515 \\ \hline


    {(d)} {\titleabbr} $-$ preprocessing & \multirow{2}{*}{\textbf{b + r}} & 95.37 & 96.27 & 73.15 & 63.42 & 0.0697 & 75.59 & 70.86 & 0.0581 \\ 

    {(e)} {\titleabbr} $-$ review filter &  & 96.50 &  97.17 & 79.60 & 70.85 & 0.0511 & 79.62 & 75.21 & 0.0458 \\ \hline
    
    {(f)} {\titleabbr} $-$ \(\Phi_c\) & \multirow{3}{*}{\textbf{b + r}} &  \multirow{3}{*}{\textbf{96.91}} & \multirow{3}{*}{\textbf{97.50}} & 81.73 & 74.78 & 0.0441 & 80.59 & 75.87 & 0.0440 \\ 
    
    {(g)} {\titleabbr} $-$ \(\mathcal{T}_k\) & &  &   & 80.27 & 73.94 & 0.0478 & 81.18 & 77.04 & 0.0420 \\ 
    
    {(h)} {\titleabbr} $-$ \(\mathcal{X}_l\)& &  &  & 80.94 & 74.79 & 0.0465 & 78.89 & 74.00 & 0.0495 \\ \hline
    
    {(i)} {\titleabbr} $-$ hierachy & \textbf{b + r} & -  & - & 63.02 & 58.98 & - & 62.34 & 48.41 & - \\ \hline

    {\titleabbr} & \textbf{b + r} & \textbf{96.91} & \textbf{97.50} & \textbf{81.89} & \textbf{77.24} & \textbf{0.0436} & \textbf{81.86} & \textbf{78.57} & \textbf{0.0401} \\ \hline

    \multicolumn{10}{r}{\scriptsize \(\Phi_c\): Label co-occurence network;~ \(\mathcal{T}_k\): Word embedding;~ \(\mathcal{X}_l\): Learnable label embedding matrix; \textbf{b}: blurb; \textbf{r}: review; ~ {\titleabbr} ($\lambda_1, \lambda_2$) is shown for {(a)}-{(c)}}


     \end{tabular}
\end{adjustbox}
\label{tab:module_ablation}
\end{table*}

\emph{(vi) Open-source LLM-based Methods:}
To comprehensively evaluate the effectiveness of {\titleabbr} against recent generative models, we consider three open-source LLM settings: zero-shot, few-shot, and supervised fine-tuning. For zero-shot and few-shot evaluation, we employ three recent large-scale 4-bit quantized instruction-tuned LLMs, namely Mistral-Small-3.2-24B-Instruct \cite{mistral7b2023}, Llama-3.3-70B-Instruct \cite{llama}, and Qwen2.5-72B-Instruct \cite{qwen3}. In the few-shot setting, we adopt dynamic 3-shot prompting, where examples are selected adaptively based on semantic relevance to the query instance instead of using a fixed demonstration pool, enabling more contextually aligned inference. For supervised fine-tuning, we restrict experiments to comparatively smaller open-weight LLMs, namely Gemma2-2B \cite{gemma2024}, Qwen3-4B \cite{qwen3}, and Mistral-7B \cite{mistral7b2023}, due to the substantially higher computational cost and memory requirements of larger models. All fine-tuned LLMs are trained using Low-Rank Adaptation (LoRA) \cite{lora}, where the LoRA rank and scaling factor are set to 64 and 32, respectively. To ensure fair comparison, all LLMs are evaluated under identical train-validation-test splits used by {\titleabbr}. As shown in Table~\ref{tab:sota_llm}, zero-shot LLMs achieved relatively limited performance, particularly on macro-level metrics, indicating difficulties in fine-grained hierarchical genre prediction without task adaptation. Few-shot prompting substantially improved performance, while fine-tuned smaller LLMs further benefited from supervised adaptation. Among all LLM baselines, Mistral-Small-3.2-24B-Instruct achieved the strongest few-shot performance, and Mistral-7B achieved the strongest fine-tuned performance. Nevertheless, {\titleabbr} consistently outperformed all LLM settings across both \emph{fiction} and \emph{non-fiction} genres. These results highlight the advantage of {\titleabbr}’s task-specific hierarchical design and review filtering strategy for structured book genre classification.

\emph{(vii) Closed-source LLM-based Methods:} 
To compare against proprietary generative models, we evaluate three recent closed-source LLMs: DeepSeek-v3-chat \cite{deepseek-v3}, Gemini-2.5-Flash \cite{gemini2_5_flash2025}, and GPT-4.1-mini \cite{openai2025gpt4.1mini}. All models are accessed through their official APIs and evaluated under both zero-shot and few-shot settings on the same evaluation instances used by {\titleabbr}. In the zero-shot setting, prediction is performed purely through instruction-based in-context learning without any task-specific fine-tuning. For the few-shot setting, we employ dynamic 3-shot prompting. As shown in Table~\ref{tab:sota_llm}, zero-shot prompting yielded moderate performance, while few-shot prompting significantly improved genre prediction. Gemini-2.5-Flash and DeepSeek-v3-chat achieved competitive few-shot results and outperformed {\titleabbr} on a few fiction metrics, but the gap on non-fiction genres remained reasonable. Overall, {\titleabbr} achieved the most consistent performance, highlighting the effectiveness of hierarchical modeling and filtered review integration.

\subsection{Ablation Studies}
\label{subsec:ablation_study}
Table~\ref{tab:module_ablation} summarizes the ablation study of {\titleabbr}. We first analyze the role of modalities (blurb, review). Removing reviews (case-(a)) (i.e., $\lambda_1$ = $\lambda_2$ = $1$) causes substantial Level-2 degradation, reducing $\mathcal{F}_{m}$ by $22.17\%$ for \emph{fiction} and $13.27\%$ for \emph{non-fiction}, highlighting the importance of reviews for fine-grained classification. In contrast, using only reviews (case-(b)) underperforms at Level-1 but improves over blurbs alone at Level-2, indicating that blurbs support coarse classification while reviews provide richer multi-label signals. Equal fusion (case-(c)) further improves performance, while the best results are achieved with $\lambda_1{=}0.3$ and $\lambda_2{=}0.7$, obtained via grid search (Section \ref{appendix:lambda_weights}), confirming the stronger contribution of reviews.

We next evaluate preprocessing and review filtering. Removing preprocessing (case-(d)) noticeably reduces Level-2 performance, while removing review filtering (case-(e)) leads to smaller but consistent degradation. For Level-2 modeling, replacing the label co-occurrence graph $\Phi_c$ with a linear classifier (case-(f)) reduces $\mathcal{F}_{m}$ by 3.28\% for \emph{fiction} and 3.42\% for \emph{non-fiction}, confirming the importance of label dependency modeling. Similarly, removing genre-aware word embeddings $\mathcal{T}_k$ (case-(g)) or learnable label embeddings $\mathcal{X}_l$ (case-(h)) further degrades performance. Finally, flattening the hierarchy into a single 57-class model (case-(i)) results in the largest performance drop, demonstrating the effectiveness of hierarchical modeling.

\begin{figure}[!b]
    \begin{minipage}{0.49\columnwidth}
    \centering
    \begin{adjustbox}{width=\textwidth}
        \begin{tikzpicture}[scale=0.60]
            \begin{axis}[
                width=1.65\textwidth,
                height=\textwidth,
                title={\scriptsize(a) Level-2: Fiction},
                title style={font=\normalsize\bfseries, yshift=-8pt},
                xlabel={\scriptsize{Filtering Threshold}},
                ymin=50, ymax=100,
                xmin=-5, xmax=105,
                xtick={0,15,30,45,60,75,90,100},
                ytick={50,60, 70, 80, 90, 100},
                ymajorgrids=true,
                grid style=dashed,
                every axis label/.append style={font=\small},
                every axis tick label/.append style={font=\small},
                xticklabel style={rotate=15, anchor=north, font=\small},
                yticklabel style={font=\small},
                legend columns=2,
                legend style={font=\scriptsize, draw=none, fill=none,
                              at={(0.2, 0.05)}, anchor=south west},
            ]
            
            \addplot[color=blue, mark=triangle]
                coordinates {(0,79.60)(15,80.37)(30,81.25)(45,81.07)
                             (60,81.49)(75,81.89)(90,80.45)(100,66.87)};
            \addlegendentry{$\mathcal{F}_{\mu}$}
 
            \addplot[color=orange, mark=*]
                coordinates {(0,92.80)(15,92.13)(30,92.72)(45,92.70)
                             (60,92.67)(75,92.80)(90,91.59)(100,86.63)};
            \addlegendentry{$\mathcal{BA}_{\mu}$}
 
            \addplot[color=red, mark=halfcircle]
                coordinates {(0,70.85)(15,72.53)(30,73.00)(45,74.02)
                             (60,76.40)(75,77.24)(90,72.50)(100,55.07)};
            \addlegendentry{$\mathcal{F}_{m}$}
 
            \addplot[color=teal, mark=square]
                coordinates {(0,91.91)(15,92.39)(30,92.29)(45,92.61)
                             (60,92.71)(75,92.49)(90,92.09)(100,88.21)};
            \addlegendentry{$\mathcal{BA}_{m}$}
            \end{axis}
        \end{tikzpicture}
    \end{adjustbox}
    \end{minipage}
    \hfill
    \begin{minipage}{0.49\columnwidth}
    \centering
    \begin{adjustbox}{width=\textwidth}
        \begin{tikzpicture}[scale=0.60]
            \begin{axis}[
                width=1.65\textwidth,
                height=\textwidth,
                title={\scriptsize (b) Level-2: Non-Fiction},
                title style={font=\normalsize\bfseries, yshift=-8pt},
                xlabel={\scriptsize{Filtering Threshold}},
                ymin=50, ymax=100,
                xmin=-5, xmax=105,
                xtick={0,15,30,45,60,75,90,100},
                ytick={50,60,70,80,90,100},
                ymajorgrids=true,
                grid style=dashed,
                every axis label/.append style={font=\small},
                every axis tick label/.append style={font=\small},
                xticklabel style={rotate=15, anchor=north, font=\small},
                yticklabel style={font=\small},
                legend columns=2,
                legend style={font=\scriptsize, draw=none, fill=none,
                              at={(0.2, 0.05)}, anchor=south west},
            ]
            \addplot[color=blue, mark=triangle]
                coordinates {(0,79.62)(15,79.56)(30,80.36)(45,80.37)
                             (60,80.85)(75,81.86)(90,79.90)(100,72.64)};
            \addlegendentry{$\mathcal{F}_{\mu}$}
 
            \addplot[color=orange, mark=*]
                coordinates {(0,93.22)(15,93.92)(30,94.05)(45,93.82)
                             (60,94.14)(75,94.00)(90,94.01)(100,92.11)};
            \addlegendentry{$\mathcal{BA}_{\mu}$}
 
            \addplot[color=red, mark=halfcircle]
                coordinates {(0,75.21)(15,74.78)(30,76.35)(45,76.39)
                             (60,76.52)(75,78.57)(90,75.06)(100,65.30)};
            \addlegendentry{$\mathcal{F}_{m}$}
 
            \addplot[color=teal, mark=square]
                coordinates {(0,93.51)(15,93.55)(30,93.11)(45,93.18)
                             (60,93.49)(75,93.23)(90,93.51)(100,91.41)};
            \addlegendentry{$\mathcal{BA}_{m}$}
            \end{axis}
        \end{tikzpicture}
    \end{adjustbox}
    \end{minipage}
    \caption{\small Effect of review filtering threshold on Level-2.} 
    \label{fig:review_filtering_threshold}
\end{figure}

\begin{figure*}[!t]
    \begin{minipage}{0.32\linewidth}
    \centering
    \begin{adjustbox}{width=\linewidth}
        \begin{tikzpicture}[scale=0.60]
            \begin{axis}[
            width=1.55\linewidth,
            height=0.9\linewidth,
            xlabel={$\lambda_1 \text{ or } \lambda_2 $},
            ymin=50, ymax=100,
            enlarge x limits={abs=0.05},
            xtick={0,0.1,0.2,0.3,0.4, 0.5, 0.6, 0.7, 0.8, 0.9, 1},
            ytick={50,60,70,80,90,100},
            legend columns=3,
            ymajorgrids=true,
            grid style=dashed,
            colormap/cool,
            every axis label/.append style={font=\large},
            every axis tick label/.append style={font=\large},
            legend style={font=\large},
            xticklabel style={rotate=60, anchor=east,font=\small},
            legend style={font=\small,draw=none, fill=none, at={(0.88, 0.22)}}
        ]
        \addplot[color=blue, mark=x] coordinates {
            (0, 87.80) (0.1,90.11)(0.2,96.64)(0.3,96.91)(0.4,96.83)(0.5,96.71)(0.6,96.84)(0.7,96.08)(0.8,96.16)(0.9,96.18)(1,94.70)
        };
        \legend{${\mathcal{F}}$};
        \addplot[color=red, mark=x] coordinates {
            (0, 63.48) (0.1,76.17)(0.2,70.49)(0.3,72.52)(0.4,72.67)(0.5,71.18)(0.6,72.71)(0.7,77.24)(0.8,73.65)(0.9,72.72)(1, 55.07)
        };
        \addlegendentry{F : ${\mathcal{F}_m}$}
        \addplot[color=teal, mark=x] coordinates {
            (0, 70.65) (0.1,74.15)(0.2,75.39)(0.3,75.70)(0.4,76.60)(0.5,75.69)(0.6,72.14)(0.7,78.57)(0.8,78.17)(0.9,76.16)(1, 65.30)
        };
        \addlegendentry{NF : ${\mathcal{F}_m}$}
        \end{axis}
        \end{tikzpicture}
    \end{adjustbox}
    \vspace{-2.25em}
    \caption*{\small{(a) Impact of $\lambda_1$ and $\lambda_2$}}
    \end{minipage}
    \hfill
    \begin{minipage}{0.32\linewidth}
    \centering
    \begin{adjustbox}{width=\linewidth}
        \begin{tikzpicture}[scale=0.60]
            \begin{axis}[
                width=1.75\linewidth,
                height=\linewidth,
                xlabel={\normalsize Genre ID},
                ymin=20, ymax=100,
                enlarge x limits={abs=0.5},
                xtick distance=50,
                xtick={1, 2, 4, 5, 6, 7, 8, 10, 11, 12, 13, 14, 15, 16, 17, 18, 19, 20, 21, 22, 23, 24, 25, 26, 27, 28, 29, 30},
                xticklabels={1, 2, 4, 5, 6, 7, 8, 10, 11, 12, 13, 14, 15, 16, 17, 18, 19, 20, 21, 22, 23, 24, 25, 26, 27, 28, 29, 30},
                ytick={20,40,60,80,100},
                legend columns=3,
                ymajorgrids=true,
                grid style=dashed,
                colormap/cool,
                every axis label/.append style={font=\scriptsize},
                xticklabel style={rotate=90, anchor=east,font=\footnotesize},
                every axis tick label/.append style={font=\small},
                legend style={font=\small,draw=none, fill=none, at={(0.9, 0.22)}}
            ]
            \addplot[color=blue, mark=triangle] coordinates {
            (1,74.62) (2,75.68) (4,100) (5,86.93) (6,86.11)(7,55.76)(8,85.71)(10,30.00)(11,40.24)(12,40.98)(13,50.00)(14,82.52)(15,65.71)(16,86.67)(17,74.36)(18,41.67)(19,83.33)(20,85.14)(21,76.97)(22,65.71)(23,72.22)(24,78.17)(25,72.34)(26,81.02)(27,54.55)(28,82.76)(29,76.20)(30,77.78)};
            \addlegendentry{$\mathcal{P}$}
            
            \addplot[color=red, mark=*] coordinates {
            (1,95.10) (2,87.50) (4,100) (5,92.68) (6,93.94)(7,100)(8,85.71)(10,100.00)(11,89.19)(12,84.03)(13,100.00)(14,94.39)(15,74.42)(16,91.76)(17,90.05)(18,100.00)(19,71.43)(20,90.52)(21,88.96)(22,95.83)(23,81.25)(24,88.80)(25,88.31)(26,81.72)(27,85.71)(28,85.71)(29,94.06)(30,63.64)};
            \addlegendentry{$\mathcal{R}$}
            
            \addplot[color=green, mark=halfcircle] coordinates {
            (1,83.62) (2,81.16) (4,100) (5,89.71) (6,89.86)(7,71.43)(8,85.71)(10,46.15)(11,55.46)(12,55.10)(13,66.67)(14,88.05)(15,69.57)(16,89.14)(17,81.46)(18,58.82)(19,76.92)(20,87.74)(21,82.53)(22,77.97)(23,76.47)(24,83.15)(25,79.53)(26,81.37)(27,66.67)(28,84.21)(29,84.19)(30,70.00)};
            \addlegendentry{$\mathcal{F}$}
            \end{axis}
        \end{tikzpicture}
    \end{adjustbox}
    \vspace{-2em}
    \caption*{\small{(b) Fiction class analysis}}
    \end{minipage}
     \hfill
    \begin{minipage}{0.32\linewidth}
    \centering
    \begin{adjustbox}{width=\linewidth}
        \begin{tikzpicture}[scale=0.60]
            \begin{axis}[
                width=1.75\linewidth,
                height=\linewidth,
                xlabel={\normalsize Genre ID},
                ymin=20, ymax=100,
                enlarge x limits={abs=0.5},
                xtick={1, 2, 3, 4, 5, 6, 7, 8, 9, 10, 11, 12, 13, 14, 15, 16, 17, 18, 19, 20, 21, 22, 23, 24, 25, 27, 28, 29, 30},
                xticklabels={1, 2, 3, 4, 5, 6, 7, 8, 9, 10, 11, 12, 13, 14, 15, 16, 17, 18, 19, 20, 21, 22, 23, 24, 25, 27, 28, 29, 30},
                ytick={20,40,60,80,100},
                legend columns=3,
                ymajorgrids=true,
                grid style=dashed,
                colormap/cool,
                every axis label/.append style={font=\scriptsize},
                every axis tick label/.append style={font=\small},
                xticklabel style={rotate=90, anchor=east,font=\footnotesize},
                legend style={font=\small,draw=none, fill=none, at={(0.80, 0.22)}}
            ]
            \addplot[color=blue, mark=triangle] coordinates {
            (1,68.00) (2,73.96) (3,81.41) (4,84.91) (5,75.51)(6,85.71)(7,73.68)(8,77.50)(9,50.00)(10,69.09)(11,71.93)(12,63.41)(13,69.62)(14,85.46)(15,80.06)(16,58.56)(17,60.96)(18,69.05)(19,47.37)(20,77.78)(21,78.18)(22,60.71)(23,67.96)(24,60.00)(25,78.15)(27,78.85)(28,87.50)(29,32.47)(30,75.64)};
            \addlegendentry{$\mathcal{P}$}
            
            \addplot[color=red, mark=*] coordinates {
            (1,94.44) (2,86.59) (3,94.81) (4,91.84) (5,92.50)(6,85.71)(7,100.00)(8,91.18)(9,80.00)(10,84.44)(11,85.42)(12,70.27)(13,94.83)(14,88.60)(15,93.73)(16,97.01)(17,86.41)(18,96.67)(19,100.00)(20,89.74)(21,90.85)(22,89.47)(23,94.59)(24,75.00)(25,93.00)(27,96.09)(28,100.00)(29,80.65)(30,95.16)};
            \addlegendentry{$\mathcal{R}$}
            
            \addplot[color=green, mark=halfcircle] coordinates {
            (1,79.07) (2,79.78) (3,87.60) (4,88.24) (5,83.15)(6,85.71)(7,84.85)(8,83.78)(9,61.54)(10,76.00)(11,78.10)(12,66.67)(13,80.29)(14,87.00)(15,86.36)(16,73.03)(17,71.49)(18,80.56)(19,64.29)(20,83.33)(21,84.04)(22,72.24)(23,79.10)(24,66.67)(25,84.93)(27,86.62)(28,93.33)(29,46.30)(30,84.29)};
            \addlegendentry{$\mathcal{F}$}
            \end{axis}
        \end{tikzpicture}
    \end{adjustbox}
    \vspace{-2em}
    \caption*{\small{(c) Non-fiction class analysis}}
    \end{minipage}
   \caption{\small Performance of {\titleabbr}: (a) Impact of $\lambda_1$, $\lambda_2$; (b)-(c) Genre-wise analysis}
    \label{fig:baselines_comparision2}
\end{figure*}


\subsection{Sensitivity to Review Filtering Threshold}
\label{subsec:review_filtering}
We analyze the effect of review filtering by varying the similarity-score percentile used in the zero-shot alignment stage. Higher percentiles apply stricter filtering, whereas lower percentiles retain more potentially noisy reviews. Level-1 results, reported in the Appendix \ref{appendix:review_filter_th_supp}, remain relatively stable across thresholds. Fig.~\ref{fig:review_filtering_threshold} shows Level-2 performance for both \emph{fiction} and \emph{non-fiction} genres. Performance improves from the highly restrictive $100^{th}$ percentile to the moderate $75^{th}$ percentile, where ${\mathcal{F}}_\mu$ and ${\mathcal{F}}_m$ achieve their best values for both genre groups. Further relaxing the threshold leads to saturated or slightly degraded performance due to noisier reviews. Overall, the results indicate that moderate semantic filtering provides the best balance between information preservation and noise reduction; hence, we use the $75^{th}$ percentile threshold in {\titleabbr}.

\subsection{Impact of Logits Weights ($\lambda_1$ and $\lambda_2$)}
\label{appendix:lambda_weights} 
We investigate the impact of $\lambda_1$ and $\lambda_2$, which control the relative contributions of blurbs and reviews-based network logits at Level-1 and Level-2, respectively. Fig.~\ref{fig:baselines_comparision2}(a) illustrates the performance of {\titleabbr} under a grid search where the weighting parameters (\(\lambda_1\) or \(\lambda_2\)) are varied from 0 to 1. Notably, \(\lambda_i=0.3\) designated that the blurb-based network contributed \(30\%\) towards the final genre prediction, while the remaining \(70\%\) was contributed by the review-based network logits. When $\lambda_1 = 0$, performance degrades substantially, indicating that user reviews alone are not sufficient for reliable Level-1 classification. As $\lambda_1$ increases from 0.1 to 0.3, performance improves sharply, suggesting that incorporating moderate information from the blurb effectively improves genre prediction at Level-1. Furthermore, for intermediate values ($0.3 \leq \lambda_1 \leq 0.9$), performance remains relatively stable, with only minor fluctuations, demonstrating robustness to moderate variations. However, when $\lambda_1$ approaches 1, performance drops noticeably, indicating that overemphasizing blurb information at the expense of user reviews negatively affects coarse-grained classification.

In contrast, at Level-2, $\lambda_2$ exhibits a significantly different trend, reflecting the increased sensitivity of fine-grained classification. When $\lambda_2$ is set to lower values, performance is relatively poor for both \emph{fiction} and \emph{non-fiction} categories. Although performance fluctuates slightly across intermediate values, we achieve the best $\mathcal{F}_m$ at $\lambda_2{=}0.7$, indicating that higher contributions from blurb are more effective for fine-grained genre discrimination at Level-2. However, further increasing $\lambda_2$ beyond this point leads to a performance decline, with a sharp drop at $\lambda_2{=}1$, suggesting that excessive reliance on blurb information disrupts balanced hierarchical learning.

\subsection{Genre-wise Performance Analysis}
\label{appendix:genrewise_analysis}
Fig.~\ref{fig:baselines_comparision2}~(b)-(c) presents the genre-wise performance of {\titleabbr} (genre IDs are detailed in Appendix~\ref {appendix:dataset}). Overall, the model achieves high recall ($\mathcal{R}$) across most genres. However, several low-frequency or overlapping genres exhibit lower precision ($\mathcal{P}$), leading to reduced $\mathcal{F}$-scores. For example, \emph{fiction} genres (IDs: 7, 10, 13, 18, 27) and \emph{non-fiction} genres (IDs: 9, 19, 29) show relatively low precision despite high recall, indicating increased false positives. Similar behavior is observed for \emph{fiction} genres (IDs: 11, 12) and \emph{non-fiction} genre (ID: 20), likely due to overlap with dominant categories. In contrast, {\titleabbr} performs strongly on genres with clearer boundaries, achieving $\mathcal{F}$-scores of $100\%$ for \emph{fiction} ID: 4 and $93.33\%$ for \emph{non-fiction} ID: 28. Overall, the results demonstrate {\titleabbr}’s ability to capture genre semantics across diverse genre distributions.



\subsection{Impact of Different Encoder} \label{appendix:encoder_selection}
{{\titleabbr} primarily relies on LM encoders to generate high-dimensional contextual embeddings from book blurbs and user reviews. As discussed earlier, the encoders are first fine-tuned in a task-specific setting (binary or multi-label classification), after which their parameters are frozen during {\titleabbr} training. To identify the most suitable encoder for {\titleabbr}, we consider two categories of recent language models: masked-based LMs and autoregressive LMs.} 
We briefly described them below.

\emph{(i) Masked-based LMs:} 
BERT \cite{bert} is a foundational masked LM containing nearly 110 million parameters, pretrained using masked language modeling and next sentence prediction objectives. Its bidirectional encoding enables strong contextual representation learning from input text. However, the large parameter size of BERT motivates more efficient variants. ALBERT \cite{ALBERT} addresses this limitation through embedding factorization and cross-layer parameter sharing, reducing memory usage with nearly \(18\times\) fewer parameters and \(1.7\times\) faster training than BERT. Similarly, DistilBERT \cite{distilBERT} employs knowledge distillation during pretraining to obtain a compressed model that retains approximately \(97\%\) of BERT’s capability while using \(40\%\) fewer parameters and enabling \(60\%\) faster inference. Beyond lightweight optimization, several models improve training strategy and representation quality. RoBERTa \cite{RoBERTa} enhances BERT by removing the next sentence prediction objective and utilizing larger batches, longer sequences, and more extensive training corpora, leading to stronger contextual representations. More recently, ModernBERT \cite{modernBERT} revisits the BERT architecture with improved tokenization and training strategies to enhance computational efficiency, robustness, and adaptability across downstream NLP tasks.

\begin{table}[!t]
\centering
\caption{Performance of {\titleabbr} with LM/LLM encoders}
    \begin{adjustbox}{width=0.48\textwidth}
        \begin{tabular}{l| c|c| c|c|c| c|c|c}
            \hline
             \multicolumn{1}{c|}{\textbf{}} & \multicolumn{2}{c|}{\multirow{2}{*}{\textbf{Level-1}}} & \multicolumn{6}{c}{\textbf{Level-2}} \\ \cline{4-9} 
            
             \multicolumn{1}{c|}{\textbf{Models}} & \multicolumn{2}{c|}{}& \multicolumn{3}{c|}{\textbf{Fiction}}& \multicolumn{3}{c}{\textbf{Non-Fiction}} \\ \cline{2-9} 
             
              & \multicolumn{1}{c|}{\textbf{${\mathcal{F}}\uparrow$}} & \textbf{${\mathcal{A}_c}\uparrow$} & \multicolumn{1}{c|}{\textbf{${\mathcal{F}}_{\mu}\uparrow$}} & \multicolumn{1}{c|}{\textbf{${\mathcal{F}}_{m}\uparrow$}} & \multicolumn{1}{c|}{\textbf{${\mathcal{HL}}\downarrow$}} & \multicolumn{1}{c|}{\textbf{${\mathcal{F}}_{\mu}\uparrow$}} & \multicolumn{1}{c|}{\textbf{${\mathcal{F}}_{m}\uparrow$}} & \textbf{${\mathcal{HL}}\downarrow$} \\ \hline 
            
            
            ALBERT \cite{ALBERT} & 87.41 & 90.42 & 68.82 & 59.97 & 0.0830 & 74.48 & 71.60 & 0.0608 \\ 

            BERT \cite{bert} & \textbf{96.91} & \textbf{97.50} & \textbf{81.89} & \textbf{77.24} & \textbf{0.0436} & \textbf{81.86} & \textbf{78.57} & \textbf{0.0401} \\ 
        
            DistilBERT  \cite{distilBERT} & 95.24 & 96.17 & 79.58 & 74.88 & 0.0486 & 80.45 & 76.05 & 0.0445 \\ 

            M-BERT \cite{modernBERT}  & 94.36 & 95.42 & 65.20 & 52.56 & 0.1043 & 60.53 & 54.75 & 0.1194 \\ 

            M-BERT-L \cite{modernBERT} & 75.50 & 81.91 & 39.82 & 75.93 & 0.1526 & 43.07 & 79.30 & 0.1652 \\

            RoBERTa  \cite{RoBERTa} & 92.43 & 93.93 & 77.95 & 69.08 & 0.0541 & 76.99 & 73.22 & 0.0542 \\ 

            RoBERTa-L \cite{RoBERTa} & 88.04 &	91.16 &	 63.83 & 82.29 & 0.0740 & 69.41 & 86.51 & 0.0526
            
            \\ \hline

             Gemma2-2B  \cite{gemma2024} & 95.58 & 94.43 &	64.83 & 51.62 & 0.0984 & 66.38 & 58.46 & 0.0901 \\

            Qwen3-4B  \cite{qwen3} & 96.43 & 95.56 & 57.26 & 44.57 & 0.1329 & 59.15 & 53.70 & 0.1206 \\

            XLNet \cite{XLNet}  & 88.92 & 91.11 & 75.01 & 64.50 & 0.0633 & 69.42 & 65.16 & 0.0806 \\

            
            \hline
        \end{tabular}
    \end{adjustbox}
    \label{tab:lm_{\titleabbr}_results}
\end{table}

\emph{(ii) Autoregressive LMs:} 
Unlike masked LMs, autoregressive LMs generate text sequentially using causal decoding objectives. XLNet \cite{XLNet} extends autoregressive pretraining through permutation-based training, combining autoregressive and autoencoding objectives for improved contextual learning, although its training and fine-tuning remain relatively complex. Recent decoder-only LMs mainly focus on improving efficiency and reasoning ability. Qwen3-4B \cite{qwen3} is a dense causal LM designed for strong multilingual and instruction-following performance. Gemma2-2B \cite{gemma2024} is a lightweight open-weight transformer that employs grouped-query attention and distillation-based training for efficient reasoning with context lengths up to \(8000\) tokens, though its smaller scale may limit representation capacity. Mistral-7B \cite{mistral7b2023} is a compact yet strong autoregressive LM that outperforms several larger models on reasoning benchmarks while supporting an \(8000\)-token context window. However, autoregressive LMs generally remain computationally expensive during inference and are not specifically designed for hierarchical multi-label classification.

Table~\ref{tab:lm_{\titleabbr}_results} presents the performance of {\titleabbr} with different LM encoders spanning both masked-based and autoregressive architectures. Overall, BERT consistently achieved the strongest and most stable performance across nearly all evaluation metrics at both hierarchical levels, justifying its selection as the primary encoder in {\titleabbr}. Its bidirectional contextual encoding enables effective modeling of long-range semantic dependencies and fine-grained lexical interactions within blurbs and reviews, which is particularly important for hierarchical genre reasoning and multi-label prediction.

Among masked-based encoders, ALBERT exhibited comparatively weaker performance, likely due to its parameter-sharing and compressed architecture reducing representational capacity for fine-grained semantic specialization. DistilBERT achieved competitive results while remaining consistently below BERT, suggesting that aggressive model compression leads to moderate degradation in semantic expressiveness. RoBERTa also underperformed relative to BERT across most settings despite stronger pretraining, indicating that larger-scale pretraining alone does not necessarily translate to improved hierarchical literary understanding. ModernBERT variants demonstrated inconsistent behavior, occasionally achieving competitive macro-level scores while exhibiting reduced micro-level stability and higher Hamming Loss, suggesting difficulties in maintaining balanced performance across overlapping genre labels.

Among autoregressive encoders, Gemma2-2B and Qwen3-4B achieved competitive Level-1 performance for coarse-grained fiction versus non-fiction discrimination, but showed substantial degradation at Level-2 during fine-grained multi-label prediction. This behavior indicates that autoregressive representations are comparatively less effective for capturing dense hierarchical label interactions and fine-grained semantic specialization required for literary genre reasoning. Overall, the results suggest that masked-based bidirectional encoders provide more robust and semantically stable representations for hierarchical genre classification. Consequently, despite a moderately higher computational cost, BERT offers the most favorable trade-off between predictive performance, semantic consistency, and hierarchical reasoning capability within {\titleabbr}.

\subsection{Impact of Different Word Embeddings}
\label{appendix:word_embedding}
To evaluate the impact of different word embedding choices on {\titleabbr}, we compare FastText \cite{fasttext}, Word2Vec \cite{w2v}, and GloVe \cite{glove}, as presented in Table \ref{tab:word_embedding_ablation}. GloVe consistently outperforms both FastText and Word2Vec, suggesting that its global co–occurrence–based representations better capture semantic regularities that align more closely with fine-grained genre distinctions at Level-2 genres modeling. 


\begin{table}[!hbt]
\centering
\caption{Performance of {\titleabbr} with different word embeddings}
    \begin{adjustbox}{width=0.48\textwidth}
        \begin{tabular}{l |cccc | cccc}
            \hline
            \textbf{Word Embedding}  & \multicolumn{4}{c|}{\textbf{Fiction (F)}} &  \multicolumn{4}{c}{\textbf{Non-Fiction (NF)}} \\ \cline{2-9}
            \textbf{Methods}  &   \(\mathcal{F}_\mu\uparrow\) & \(\mathcal{BA}_\mu\uparrow\) & \(\mathcal{F}_m\uparrow\) & \(\mathcal{BA}_m\uparrow\) & \(\mathcal{F}_\mu\uparrow\) & \(\mathcal{BA}_\mu\uparrow\) & \(\mathcal{F}_m\uparrow\) & \(\mathcal{BA}_m\uparrow\) \\ \hline
            FastText \cite{fasttext} & 78.77 & 92.14 & 70.95 & 92.13 & 76.72 & 93.62 & 71.81 & 92.93\\
            Word2Vec \cite{NIPS2013_word2vec} & 79.25 & 92.48 & 70.51 & 92.44 & 79.86 & 93.74 & 75.31 & 92.53\\
            Glove \cite{glove} & \textbf{81.89} & \textbf{92.80} & \textbf{77.24} & \textbf{92.49} & \textbf{81.86} & \textbf{94.00} & \textbf{78.57} & \textbf{93.23} \\ \hline
        \end{tabular}
    \end{adjustbox}
    \label{tab:word_embedding_ablation}
\end{table}




\begin{figure*}[!t]
    \centering
    \includegraphics[width=0.75\linewidth]{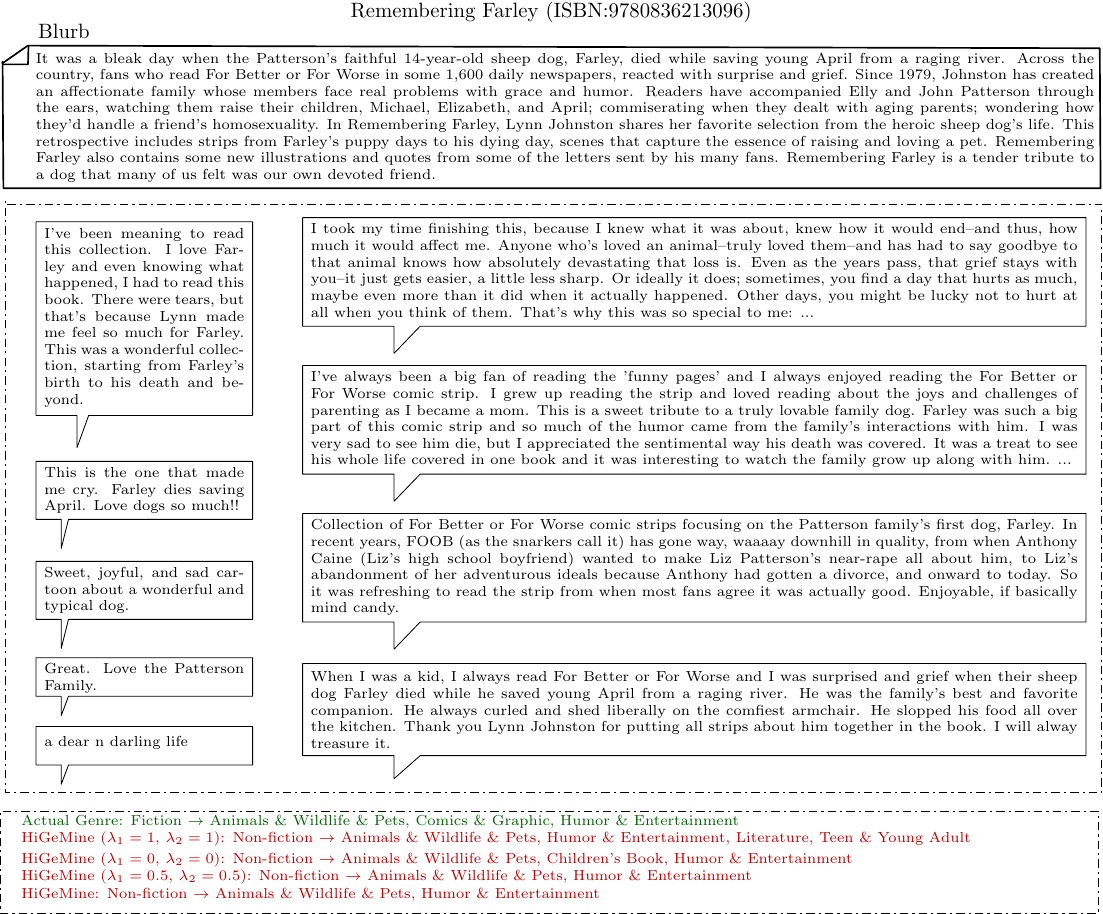}
    \caption{Qualitative analysis for misprediction due to Level-1 classification}
    \label{fig:misprediction_due_to_level_1_classification}
\end{figure*}

\subsection{Qualitative Analysis} \label{appendix:qualitative_analysis}
In this sub-section, we provide the qualitative analysis of {\titleabbr}, exploiting the following details.

\begin{figure*}
    \centering
    \includegraphics[width=0.8\linewidth]{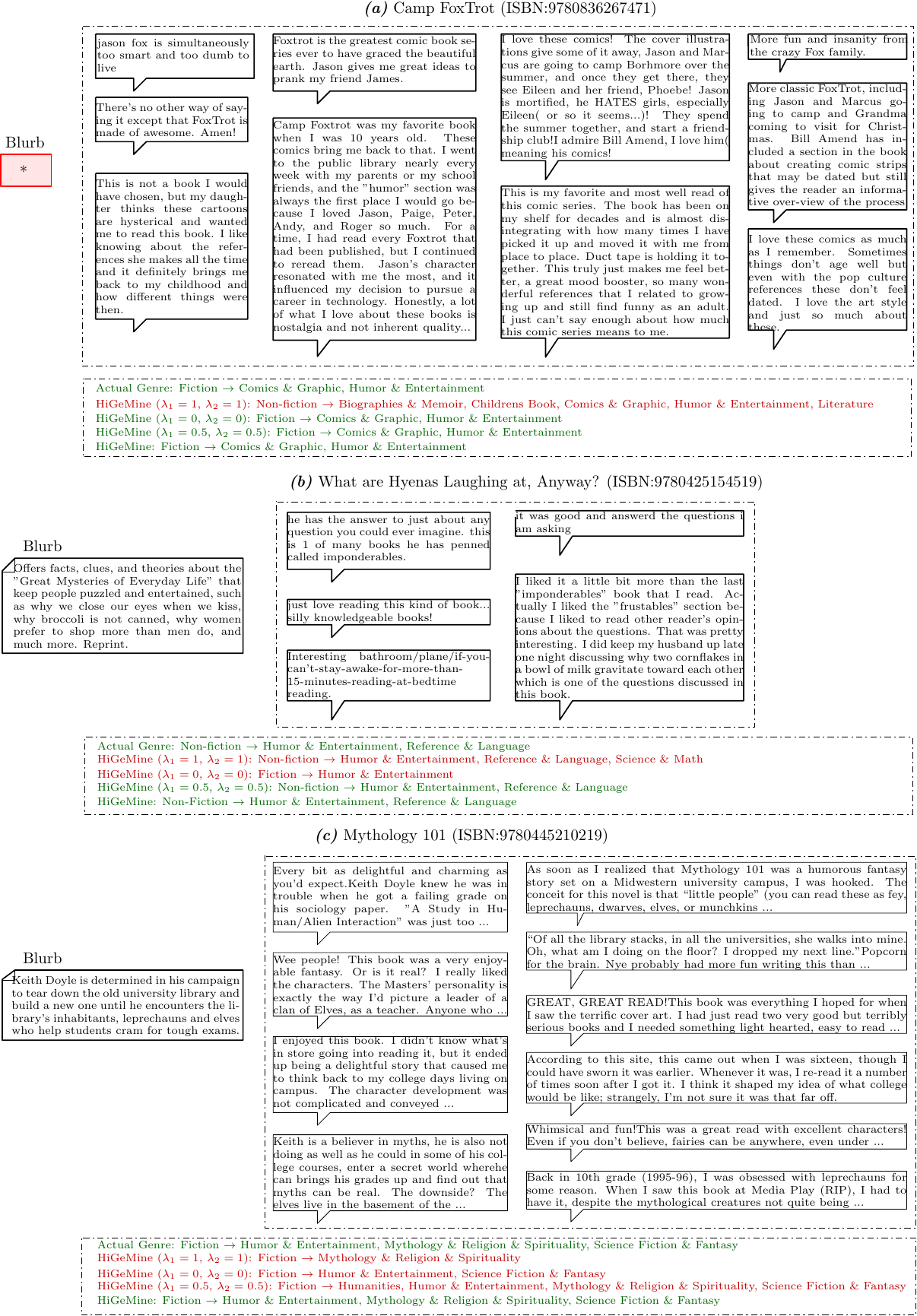}
    \caption{Qualitative analysis of modality ablation:
    \textit{\textbf{(a)}} Scarce blurbs lead to incorrect predictions, while reviews alone or combined with blurbs yield correct genres;
    \textit{\textbf{(b)}} Informative blurbs and scarce reviews may still mislead individually, but their combination enables correct classification;
    \textit{\textbf{(c)}} Even equal contributions from both modalities may fail, whereas {\titleabbr} adaptively balances them to predict accurately. Note that a few reviews are truncated due to over length.}
    \label{fig:modality_qualitative_figure}
\end{figure*}

\emph{(i) Modality Analysis:}
\label{appendix:qual_anal_cls} 
Fig.~\ref{fig:modality_qualitative_figure} shows the qualitative analysis of {\titleabbr} classification ability through three sample scenarios (\emph{\textbf{a}}, \emph{\textbf{b}}, \emph{\textbf{c}}) involving book blurbs and reviews. The analysis compares predictions across four model configurations with varying modality weightings ($\lambda_1$, $\lambda_2$): {\titleabbr} (1, 1), {\titleabbr} (0, 0), {\titleabbr} (0.5, 0.5), and standard {\titleabbr}.

\noindent \---- \textit{\textbf{Sample (a)}: Information-spare blurb with comprehensive reviews:} {\titleabbr} (1, 1) failed to correctly classify genres due to over-reliance on the uninformative blurb, while other configurations succeeded by effectively leveraging review content. This highlights the model's adaptability when the blurb is inadequate.

\noindent \---- \textit{\textbf{Sample (b)}: Informative blurb with limited reviews:} {\titleabbr} (0, 0) was completely misclassified as it only used limited reviews, and {\titleabbr} (1, 1) achieved partial accuracy. Other configurations demonstrated full accuracy through the integration of both modalities.

\noindent \---- \textit{\textbf{Sample (c)}: Sufficient content in the blurb and reviews:} Standard {\titleabbr} achieved perfect genre prediction by complementary from both blurb and reviews. At the same time, other configurations showed reduced performance due to suboptimal modality weighting.

This demonstrates standard {\titleabbr}'s superior cross-modal integration capabilities compared to its variants, particularly in scenarios requiring dynamic adaptation to variable information quality across blurb and reviews.

\begin{figure*}[!t]
    \centering
    \includegraphics[width=0.75\linewidth]{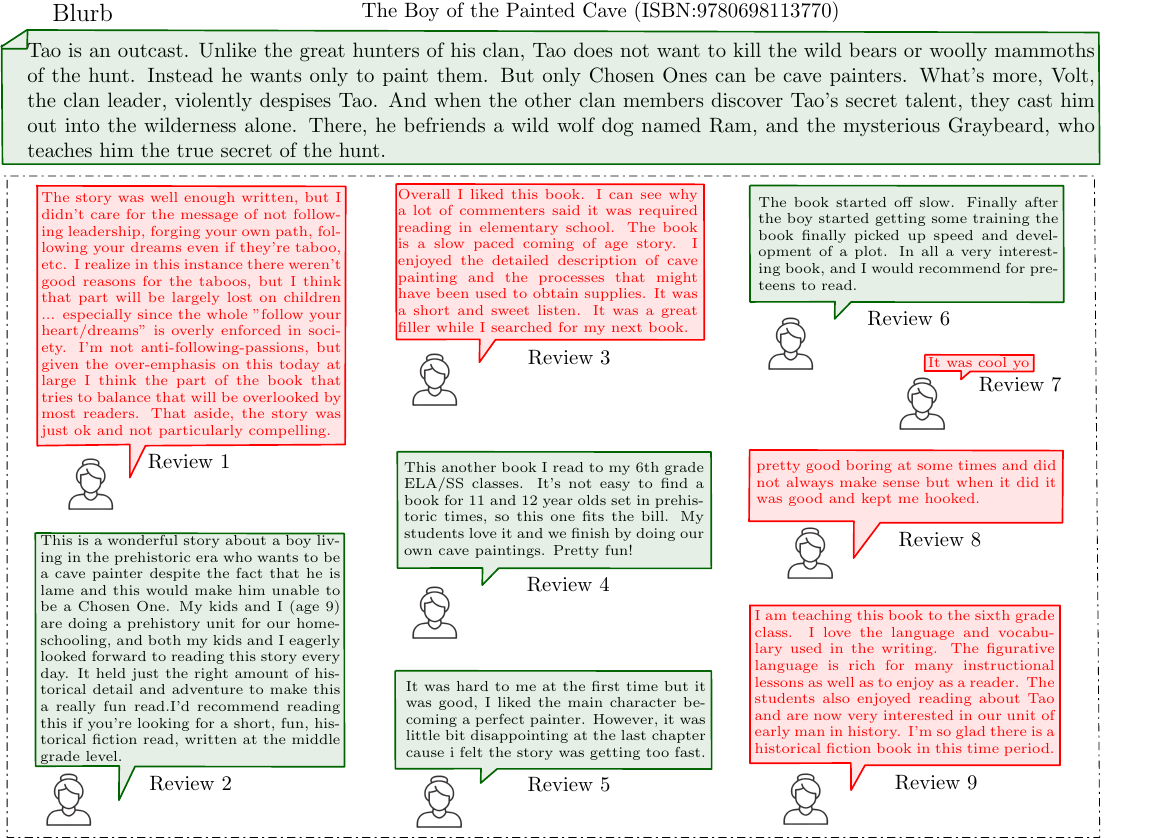}
    \caption{Qualitative analysis for review filtering for a sample blurb. Filtered-in and filtered-out reviews are highlighted in \colorbox{green!20}{green}, and \colorbox{red!20}{red} color, respectively.}
    \label{fig:review_filtering_qualtitative_example}
\end{figure*}

\emph{(ii) Review Filtering:}
Fig.~\ref{fig:review_filtering_qualtitative_example}
presents a qualitative analysis demonstrating the efficacy of review filtering for genre classification. The \colorbox{green!20}{green} highlighted reviews represent those selected by our filtering mechanism based on their relevance to informational content. These selected reviews contain substantial genre-indicative information about the book. For instance, Fig.~\ref{fig:review_filtering_qualtitative_example} provides genre through phrases, \emph{Review 2}: \emph{``a boy living in the prehistoric era who wants to be a cave painter"}, \emph{``middle grade level"}, \emph{``historical detail and adventure"}, and characterizes it as a short, fun, historical fiction read. Such phrases offer information for genre classification, specifically indicating alignment with \textit{fiction}, \textit{children's book}, and \textit{history} categories. 


In contrast, reviews that were not selected exhibit significant information sparsity, exemplified by the review, \emph{Review 7}: \emph{``It was cool yo"}, which does not provide genre-relevant content. Further, the filtering process systematically excludes reviews that contain potentially misleading information, including personal opinions, \emph{Review 3}: \emph{``Overall I liked this book"} or offering generic reviews without genre-specific content, \emph{Review 8}: \emph{``pretty good boring at some times"}. Such reviews focus primarily on the reader experience rather than the characteristics of the content.

This review filtering approach ensures that {\titleabbr} focuses on high-quality, noise-free, genre-relevant data, thereby boosting accuracy and performance.


\emph{(iii) Misprediction Analysis:}
We observe certain Level-2 mispredictions arise from error propagation originating at Level-1. Fig.~\ref{fig:misprediction_due_to_level_1_classification} presents a representative example of this error propagation. As Level-2 inference is explicitly conditioned on the coarse-grained genre prediction, an incorrect Level-1 assignment restricts the model to an incompatible subgenre set, making accurate fine-grained classification infeasible. Detailed misprediction analysis is provided in the Appendix \ref{appendix:misprediction_analysis}. 

\subsection{Computational Efficiency Analysis}
\label{appendix:computational_efficiency}
Table~\ref{tab:computational_cost} presents a comparative analysis of computational efficiency in terms of FLOPs (trillions) and inference latency (seconds). {\titleabbr} (0.1345 TFLOPs, 0.0659s) incurs the highest computational cost among non-LLM baselines, attributable to its semantic-role-separated graph propagation and structured genre interaction modeling. Nevertheless, this overhead is offset by substantial gains in hierarchical genre prediction performance (refer Tables~\ref{tab:baseline_results_main}-\ref{tab:sota_llm}), reflecting an effective trade-off between modeling capacity and cost. Critically, {\titleabbr} remains far more efficient than large-scale LLMs: approximately 13.6× fewer FLOPs than Mistral-7B and over 37× lower latency than GPT-4.1-mini, while achieving superior or competitive performance. The absolute latency of 0.0659s confirms that {\titleabbr}'s hierarchical multimodal architecture remains practically deployable without sacrificing predictive performance.

\begin{table}[!t]
\centering
\caption{Computational efficiency}
    \begin{adjustbox}{width=0.48\textwidth}
        \begin{tabular}{l|l|c|c}
            \hline
           \textbf{Categor}y & \textbf{ Models}          & \textbf{TeraFLOPS} & \textbf{Latency (second)} \\ \hline
           \multirow{2}{*}{SOTA} & N.-Flores et al. \cite{desc_spanish_ieee_access_2023} & 0.0217    & 0.0171      \\ 
                                 & Ullah et al. \cite{desc_bangla_2023_iccit} & 0.0220 & 0.0219 \\ \hline
                                 
           \multirow{2}{*}{MLTC} & BiG \cite{BiG_ESA_2025} & 0.0435 & 0.0024 \\
                                 & UCLAF \cite{UCLAF_IJCAI_2025} & 0.0787 & 0.0072 \\ \hline
                                 
           \multirow{2}{*}{HTC} & HILL \cite{hill}  & 0.0262   & 0.0307      \\ 
                                & L-SIMCL \cite{LSIMCL_IPM_2026} & 0.0445 & 0.0041  \\ \hline

            \multirow{3}{*}{Graph-based} & BertGCN \cite{bertgcn_ijcnlp_2021} & 0.1284 & 0.0350 \\ 
                                         & CorGCN \cite{CorGCN_KDD25} & 3.85e-6 & 5.7e-5   \\ 
                                         & HiGATA \cite{HiGATA_IJCNN_2025} & 0.0437  &  0.0115 \\  \hline
           LM & DistilBERT   \cite{distilBERT}    & 0.0612     & 0.0425      \\  
           Open Source LLM & Mistral-7B  \cite{mistral7b2023}     & 1.8300       & 0.3511 \\ 
           Close source LLM & GPT-4.1-mini  \cite{openai2025gpt4.1mini}  & -  & 2.4876 \\ \hline
           \multicolumn{2}{c|}{{\titleabbr}} & 0.1345    & 0.0659      \\ \hline
        \end{tabular}
    \end{adjustbox}
\label{tab:computational_cost}
\end{table}


To provide a comprehensive understanding of {\titleabbr}, the Appendix \ref{appendix:dataset} provides dataset details, \ref{appendix:implementation_and_hyperparameter_details} contains hyper-parameter configurations, \ref{appendix:review_filter_th_supp} includes review filtering sensitivity, \ref{appendix:misprediction_analysis} extends misprediction and error analysis, and \ref{limitations} presents limitations.


\section{Conclusion} 


We presented {\titleabbr}, a hierarchical book organization framework that
integrates publisher-authored blurbs with semantically refined crowdsourced
reviews to derive structured genre metadata. Blurb-guided semantic refinement
suppresses noisy and genre-irrelevant review evidence, while dual-path graph
convolutions preserve complementary blurb--review semantics and model
inter-genre dependencies for coarse-to-fine multi-label classification.
Experiments on our proposed hierarchical dataset demonstrate that {\titleabbr}
outperforms flat, hierarchical, graph-based, transformer, and LLM-based
baselines, supporting its effectiveness for reliable book organization.
The resulting fine-grained semantic metadata provides a scalable basis for
textual learning-resource discovery in digital learning environments. Future
work will investigate multimodal evidence, deeper genre hierarchies, and
downstream learning-resource retrieval and recommendation.




\bibliographystyle{IEEEtran} 
\bibliography{custom}






\section*{Supplementary Appendix}
Appendices are provided at \url{https://github.com/csksuraj17/HiGeMine_supp}

\end{document}